\documentclass[11pt,a4paper]{article}
\usepackage[margin=2.5cm,bottom=2.5cm]{geometry}
\usepackage[latin1]{inputenc}
\usepackage{amsmath}
\usepackage{amsfonts}
\usepackage{amssymb}
\usepackage{graphicx}

\usepackage{graphicx}
\usepackage{amsfonts,dsfont}
\usepackage[round]{natbib}
\usepackage{enumerate}
\usepackage{subfigure} 

\usepackage{comment}
\usepackage{rotating}
\usepackage{amssymb} 
\usepackage{amsfonts}
\usepackage{mathtools}
\usepackage{float}
\usepackage{amsfonts,dsfont}
\usepackage{hyperref}
\usepackage{xcolor}
\hypersetup{
	colorlinks=true,
	linkcolor=blue,
	citecolor = blue,
	urlcolor=blue,
}
\usepackage{rotating}
\usepackage{lscape}
\usepackage{afterpage}
\usepackage{flafter}
\usepackage{fancyhdr}
\usepackage[linesnumbered,ruled,vlined]{algorithm2e} 
\newcommand{\iid}{\stackrel{iid}{\sim}}

\graphicspath{{Images/}}

\begin{document}
	\title{Bayesian estimation of probabilistic sensitivity measures}
    \author{Isadora Antoniano-Villalobos \thanks{Department of Environmental Sciences, Informatics and Statistics, Ca' Foscari University of Venice, Italy and Bocconi Institute for Data Science and Analytics (BIDSA), Bocconi University, Milan, Italy.    }  	
    	\and
        Emanuele Borgonovo \thanks{Department of Decision Sciences and BIDSA, Bocconi University,	Milan, Italy. } 
        \and
        Xuefei Lu \thanks{Department of Energy, Politecnico di Milano and Department of Decision Sciences, Bocconi University,	Milan, Italy.}
    }

\date{}
\maketitle

\begin{abstract}
Computer experiments are becoming increasingly important in scientific investigations. In the presence of uncertainty, analysts employ probabilistic sensitivity methods to identify the key-drivers of change in the quantities of interest. Simulation complexity, large dimensionality and long running times may force analysts to make statistical inference at small sample sizes. Methods designed to estimate probabilistic sensitivity measures at relatively low computational costs are attracting increasing interest. We propose a fully Bayesian approach to the estimation of probabilistic sensitivity measures based on a one-sample design. We discuss, first, new estimators based on placing piecewise constant priors on the conditional distributions of the output given each input, by partitioning the input space. We then present two alternatives, based on Bayesian non-parametric density estimation, which bypass the need for predefined partitions. In all cases, the Bayesian paradigm guarantees the quantification of uncertainty in the estimation process through the posterior distribution over the sensitivity measures, without requiring additional simulator evaluations. The performance of the proposed methods is compared to that of traditional point estimators in a series of numerical experiments comprising synthetic but challenging simulators, as well as a realistic application.\\
\emph{An Updated Version of the Manuscript is Forthcoming in Statistics and Computing.}
\end{abstract}

\section{Introduction}\label{intro}
The use of computer simulations is becoming increasingly important in broad areas of science  \citep{LinBing10,Wong2017635}. High-fidelity mathematical models allow analysts to perform virtual (or \emph{in silico}) experiments on complex natural or societal phenomena of interest \citep[see][among others]{Smitetal12}. Predictions are often used to support policy-making. However, the level of sophistication of the models is often too high for analytical solutions to be available. In these cases, the only way to obtain a quantitative solution may be to encode complex mathematical equations in a computer software; so that the input-output mapping remains a black-box to the analyst. It then becomes important to carefully design and execute the computer experiment. The design and analysis of computer experiments (DACE) has entered the statistical literature with the seminal work of \cite{Sacks1989} (see also the monographs of \cite{Santner2003,Kleijnen2008}). Since then, researchers have studied the creation of space-filling designs \citep{PronMuel12,He17}, the calibration of computer codes with real data \citep{TuoWu15}, their emulation \citep{Marrel2012833}, the quantification of uncertainty in their output \citep{Oakley02,Ghanem2016} and their sensitivity analysis \citep{OaklOhag04,BorgTara12JRSSB}. These areas are intertwined. A given design may allow, for instance, not only an uncertainty quantification, but also the creation of an emulator and the analysis of sensitivity. 

Probabilistic (or global) sensitivity measures are an indispensable complement of uncertainty quantification, as they highlight which areas should be given priority when planning data collection or further modelling efforts. International agencies such as the US Environmental Protection Agency \citep{USEPA2009} or the British National Institute for Health Care Excellence \citep{NICE2013} and the \cite{ERC09} have issued guidelines recommending the use of probabilistic sensitivity analysis methods as the gold standard for ensuring reliability and transparency when using the output of a computer code for decision-making under uncertainty. Over the years, several probabilistic sensitivity measures have been proposed. Different measures enjoy alternative properties making them preferable in different contexts and for different purposes. We recall regression-based \citep{Helton2009},  variance-based \citep{saltelli_jasa_2002,RugaGilq18} and moment-independent measures \citep{BorgTara12JRSSB}, all of which offer alternative ways to quantify the degree of statistical dependence between the simulator inputs and the output. A transversal issue in realistic applications is that analytical expressions of these measures are unavailable  and analysts must resort to estimation. This, however, is a challenging task, especially for simulators with a high number of inputs (the \textit{curse of dimensionality}) or with long running times (high \emph{computational cost}). 

Recent results \citep{StroOaki12JRSSC,StroOakl13} evidence the \emph{one-sample} (or \emph{given-data}) approach as an attractive design, which allows analysts to estimate global sensitivity measures from a single \textit{probabilistic sensitivity analysis sample}, i.e., a sample generated for propagating uncertainty in the simulator. Thus, a one-sample approach has a nominal cost equal to  the sample size and independent of the number of inputs, a feature that potentially reduces the impact of the curse of dimensionality. Related works such as \cite{StroOaki12JRSSC,StroOakl13,PlisBorg13, StroOakl14,Strong2015570} and \cite{BorgHazePlish15} provide advances on theoretical and numerical aspects of the methodology, while at the same time, evidencing some open research questions. One-sample estimation procedures are closely related to scatter-plot smoothing, where partitioning of the covariate space plays a central role  \citep{hastie1990generalized}. \cite{StroOakl13} show that the choice of partition size affects estimation, especially when the sample size is small (see Fig. 1 of \citep[p. 759]{StroOakl13}) . In the literature, some heuristics for determining a partition selection strategy which is optimal in some sense have been proposed, but finding a universally valid heuristic seems out of reach  (see  Appendix \ref{Sec:AppHeuristic} for numerical experiments illustrating this issue). Moreover,  the literature is concerned with point estimators and uncertainty regarding the estimated value of a sensitivity measure is not an intrinsic part of the analysis. Because most one-sample estimators are consistent (in the frequentist sense), an accurate estimation of the error is often overlooked. However, especially at small sample sizes, it is essential for transparency that interval estimates become part of result communication (see \cite{JanoNodePrie14} among others).    

We propose to enrich the one-sample design through the use of Bayesian non-parametric (BNP) methods, aiming to reduce and even eliminate the partition selection problem, while making uncertainty in the estimates a natural ingredient. First, we extend the partition approach, using Bayesian non-parametric models to augment the output sample within each partition set, by adequately generating additional synthetic data according to two alternative schemes. The Bayesian intuition supporting these designs can be interpreted as setting a prior on the conditional distribution of the output, given that the input realization falls within a given set of the partition. We build estimators based on this intuition for variance-based, density (pdf)-based and cumulative distribution function (cdf)-based global sensitivity measures. We compare the results with given-data estimators currently in use at low sample sizes, through numerical experiments. The results show that  our estimators recover the correct ranking of the inputs, while providing an appropriate quantification of the estimation uncertainty. However, the results may be strongly influenced by the partition choice. Therefore, we investigate two additional classes of estimators based on Bayesian non-parametric joint and conditional density estimation methods. These estimators eliminate the partition selection problem and, at the same time, enable error quantification. Finally, we discuss the application of all the new estimators to the global sensitivity analysis of the benchmark computer code known as LevelE \citep{saltelli_jasa_2002}.  Results show that the estimators correctly identify the key drivers of uncertainty at sample sizes lower than the ones used in previous literature. Additionally, the analyst obtains a quantification of the uncertainty in the estimates in the form of a posterior distribution which can be used to determine whether the available sample is large enough to make robust conclusions about the simulator input ranking.

The reminder of the paper is organized as follows. We begin in Section \ref{S:SensitivityMeasures} by introducing the framework of global sensitivity analysis and the one-sample estimation approach for probabilistic sensitivity measures. Section \ref{sec:BNPCDF} combines Bayesian non-parametric methods and the one-sample approach to create two new partition-dependent estimators. Section \ref{sec:BNPPDF} derives two classes of Bayesian partition-free estimators by adopting Bayesian a non-parametric density estimation approach. Section \ref{sec:levelE} presents numerical results for the LevelE code. Section \ref{sec:conclusions} offers discussion and conclusions.  The appendices illustrate the algorithms of the proposed estimators and discuss additional implementation details. \\

\section{Probabilistic sensitivity analysis of computer experiments}\label{S:SensitivityMeasures}
Formally, the sensitivity analysis framework considers a multivariate mapping $ g: \mathcal{X} \mapsto \mathcal{Y} $ with input space $\mathcal{X}\subseteq \mathbb{R}^k $ and output space $\mathcal{Y} \subseteq \mathbb{R}^d$, denoted as $\mathbf{y}=g(\mathbf{x})+\epsilon(\mathbf{x})$ in its more general form. In the DACE set-up, $g$ represents a set of operations performed by a computer code which processes a set $\mathbf{x}$ of inputs, resulting in a set $\mathbf{y}$ of outputs of interest. The term $ \epsilon(\mathbf{x}) $ represents a zero-mean error term, which is present when the simulator response is stochastic. For simplicity, we focus on deterministic univariate responses, with $ \epsilon(\mathbf{x})\equiv 0$  and $ d=1 $. When information is not sufficient to fix the values of the inputs, it is common to \textquotedblleft assume to have information about the factors'	probability distribution, either joint or marginal, with or without correlation, and that this knowledge comes from measurements, estimates, expert opinion...\textquotedblright \citep[][p. 704]{saltelli_jasa_2002}. We denote the input probability space by $(\mathcal{X},\mathcal{B(X)},\mathbb{P}_{\mathbf{X}})$, where $\mathbb{P}_{\mathbf{X}}$ represents the joint probability measure of $\mathbf{X}=(X^1,\ldots,X^k)$, assumed known. Similarly, $(\mathcal{Y},\mathcal{B(\mathcal{Y})},\mathbb{P}_Y)$ denotes the output probability space, where $\mathbb{P}_Y$ represents the distribution of $ Y $ induced by $\mathbb{P}_{\mathbf{X}}$ through $ g $. 

It has been recently shown that several probabilistic sensitivity measures frequently used in practice can be expressed as expectations of measures of discrepancy between  $\mathbb{P}_{Y}$ and $ \mathbb{P}_{Y|X^i}$. In particular, we focus on probabilistic sensitivity measures of the form: 
\begin{equation}
	\xi _{i}:=\mathbb{E}[\zeta (\mathbb{P}_{Y},\mathbb{P}_{Y|X^i})]
	\label{eq_xi}
\end{equation}%
where the expectation is calculated with respect to the marginal distribution of $X^{i}$ and $ \zeta$ is a pre-metric on the space of probability measures over $\mathcal{Y}$, and $ \xi_i $ is called the \emph{probabilistic sensitivity measure} of $X^i$ with \emph{inner operator} $ \zeta $ \citep{BorgTara12JRSSB}.

\begin{table*}
	\centering
	\caption{Some probabilistic sensitivity measures}
	\label{table_SenMeas}       
	\begin{tabular}{lll}
		\hline\noalign{\smallskip}
		Measure & $\zeta (\mathbb{P}_{Y},\mathbb{P}_{Y|X^i})$ & $\xi_i$  \\
		\noalign{\smallskip}\hline\noalign{\smallskip}
		$\eta_i$& $(\mathbb{E}[Y|X^i]-\mathbb{E}[Y])^2/\mathbb{V}[Y]$& $\mathbb{V}[\mathbb{E}(Y|X^i)]/\mathbb{V}[Y]$ \\
		$\delta_i$ & $\dfrac{1}{2}\int_{\mathcal{Y}}|f_{Y|X^i}(y|X^i)-f_{Y}(y)|\text{d}y$ &
		$\dfrac{1}{2}\mathbb{E}\left[ \int_{\mathcal{Y}}|f_{Y|X^i}(y|X^i)-f_{Y}(y)|\text{d}y\right]$ \\ 
		$\beta_i$ & $\sup_{y\in\mathcal{Y}} \left\vert F_{Y|X^i}(y|X^i)-F_Y(y) \right\vert$ & $\mathbb{E}\left[\sup_{y\in\mathcal{Y}} \left\vert F_{Y|X^i}(y|X^i)-F_Y(y) \right\vert \right]$\\
		\noalign{\smallskip}\hline
	\end{tabular}
\end{table*}

Table \ref{table_SenMeas} reports three probabilistic sensitivity measures encompassed by this construction, namely, the \emph{variance-based} sensitivity measure ($ \eta_i $), the \emph{density-based} $ \delta$-importance measure ($\delta_i$) and the \emph{cdf-based} $\beta$-importance measure ($ \beta_i $) \citep{Pear05,saltelli_jasa_2002,OaklOhag04}. The quantity $ \eta_i $ represents the expected fractional reduction in the simulator output variance attained by fixing $X^i  $ and it is equal to the popular first-order variance-based sensitivity index \citep{ImanHora90,Sobo93} when the simulator inputs are mutually independent \cite{saltelli_jasa_2002,LiuOwen06}. 

Most global sensitivity measures are used by analysts to quantify the strength of the statistical dependence of $ Y $ on $ X^i $. It has been noted that first-order variance-based sensitivity measures do not possess the \emph{nullity-implies-independence property}, known also as R\'enyi's's postulate D for measures of statistical dependence \citep{Renyi1959}. In simple terms, a null value of $ \eta_i $ does not imply that $ Y $ is independent of $ X^i $. This has led to the introduction and study of sensitivity measures that satisfy such postulate and the class of \emph{distribution-based} (or \emph{moment-independent}) sensitivity measures is attracting increasing attention in the literature \citep[see e.g.][]{GambKleiLagn15,DaVe14,Rahm14}. As representatives, we focus on the sensitivity measures $ \delta_i $ and $ \beta_i $ (Table \ref{table_SenMeas}), which quantify the expected separation between $ \mathbb{P}_{Y} $ and $ \mathbb{P}_{Y|X^i} $ through the $ L^1$-norm between densities and the Kolmogorov-Smirnov distance between cumulative distribution functions, respectively. Besides complying with Renyi's postulate D, these probabilistic sensitivity measures are also invariant to monotonic transformations -- a property that accelerates convergence in numerical estimation \citep[for further details see e.g.][]{BorgTara12JRSSB}.

As mentioned in the introduction, analytical expressions for these and other popular sensitivity measures are not available in most realistic applications, and their estimation is a prolific subject of research. We now discuss some relevant aspects of numerical estimation in the next section. 

\subsection{One-sample approach to the estimation of probabilistic sensitivity measures}\label{subsec:onesample}

The estimation of global sensitivity measures is a challenging task and the availability of efficient designs is crucial in realistic applications. The number of simulator evaluations necessary to estimate sensitivity measures encompassed by Eq. \eqref{eq_xi} for a simulator with $k$ simulator inputs, using a brute-force approach, would be of the order of $ C=kn^2 $ simulator runs, where $ n $ denotes the sample size required for Monte Carlo uncertainty quantification. The design becomes rapidly infeasible. For instance, if $ k=20 $ and $ n=1,000 $, the $ C=20,000,000 $ simulator runs would require a prohibitive computational effort for most complex computer codes used in practice. However, notable advances in the literature have contributed in abating this computational burden \citep[see][for reviews]{TissPrie15,janon2014asymptotic}. \cite{Saltelli_cpc_2002}, for instance, achieved the estimation of variance-based sensitivity measures at a cost of $C= n(k+2) $ simulator runs, while the $FAST$ method of \cite{Saltelli1999} achieves a cost of order $C= nk $. 

Recently, efforts have been made towards an estimation cost independent of the number of simulator inputs, $k$. \cite{StroOakl13}, \cite{StroOakl14} and \cite{Strong2015570} show that value-of-information measures can be estimated from a single probabilistic sensitivity analysis sample, $\{(\mathbf{x}_j,y_j):j=1,\ldots n\}$,  i.e., from the Monte Carlo sample generated for uncertainty quantification, thus lowering the computational cost to $ C=n $ simulator runs. \cite{Roehlig2009}  and \cite{StroOaki12JRSSC}  obtain similar results for first-order variance-based sensitivity measures and \cite{PlisBorg13} extend the intuition to density-based measures. The approaches proposed in these works receive the common name of \emph{one-sample} or \emph{given-data} estimation methods. 

One-sample methods can be seen as generalizations of the intuition developed for estimating the correlation ratio \citep{Pear05}. If $ X^i $ is a discrete random variable then, an input-output sample of (sufficiently large) size $n$ contains repeated observations of $ Y=g(X^i,X^{-i}) $,  for each fixed value $ X^i=x^i $, while the other factors, $ X^{-i} = (X^1, \dots, X^{i-1}, X^{i+1},\dots, X^{k}) $, remain random. This allows the estimation of $ \zeta(\mathbb{P}_Y,\mathbb{P}_{Y|X^i=x^i}) $ directly from a sample of size $ n $. For a continuous $X^i$, a similar result may be achieved by partitioning the support $\mathcal{X}^{i}$ of $ X^i $ into $ M $ bins $\{\mathcal{X}^i_m\}_{m=1}^M$. The point condition $(X^i=x^i)$ is then replaced by the bin condition $(X^i\in\mathcal{X}^i_m)$. Then, for any sensitivity measure encompassed by Eq. \eqref{eq_xi}, a one-sample estimator is given by \citep{BorgHazePlish15}:
\begin{equation}
	\widehat{\xi }_{i}=\sum_{m=1}^{M}\mathbb{P}_{X^i}(\mathcal{X}^i_m)\,\hat{\zeta}^{i}_m, 
	\label{eq_xiOS}
\end{equation} 
where $\hat{\zeta}^{i}_m$ may be any estimator of $\zeta(\mathbb{P}_{Y},
\mathbb{P}_{Y|X^i\in \mathcal{X}^i_m} )$. Note that by using equiprobable partition sets, $\mathbb{P}_{X^i}(\mathcal{X}^i_m)$ should reduce to $1/M$. In practice, this partition probability is estimated by the sample proportion,  $n^i_m/n$, where $ n^i_m $ denotes the number of realizations for which the $i$-th input falls within the $m$-th partition set of its support. \cite{BorgHazePlish15} (Theorem 2) show that, under mild conditions on the inner operator $\zeta$,  a consistent version of the estimator in Eq. \eqref{eq_xiOS} can be obtained, if the size $M$ of the partition is chosen as a monotonically increasing function of the sample size $n$, such that $\lim\limits_{n\rightarrow \infty} \frac{n}{M(n)} = \infty$.

Indeed, the most popular one-sample estimator  of  $\eta_i$ relies on a plug-in estimator of the inner statistic, based on the output sample mean and variance, $\bar{y}$ and $s^2_y$ respectively, to estimate the marginal mean and variance of $Y$. The within cluster sample mean $\bar{y}^i_m = \dfrac{1}{n^i_m}
\sum_{y\in \mathbf{y}^i_m} y$ with $\mathbf{y}^{i}_{m}=\{y_{j}:x_{j}^{i}\in \mathcal{X}^i_m, j=1,2,...,n\}$ is used to estimate the conditional mean of $Y|X^i\in \mathcal{X}^i_m$. The final expression \citep [see e.g.][]{StroOaki12JRSSC} takes the form of Eq. \eqref{eq_xiOS} with:
\begin{equation}\label{eq_etaOS}
	\widehat{\eta}_i^\star = \sum_{m=1}^{M}\frac{n^i_m}{n} \frac{(\bar{y}^i_m - \bar{y})^2}{s^2_y}.
\end{equation}
The one-sample estimator for the $\delta-$importance introduced by \cite{PlisBorg13} can be written as:
\begin{equation}\label{eq_deltaPDF}
	\widehat{\delta}_i^{\star} = \sum^{M}_{m=1} \frac{n^i_m}{n} \int_{\mathcal{Y}}|\hat f^{\star}_{Y}(y)-\hat f^i_m (y)|\text{d}y,
\end{equation}
where $ \hat{f}^{\star}_{Y} $ and $ \hat f^i_m  $ denote kernel-smoothed histograms of the full output vector $\mathbf{y} = (y_1,\ldots y_n) $ and the within cluster output vector $\mathbf{y}^{i}_{m}$, respectively. The authors propose a quadrature method for the numerical integration required by the $ L^1$-norm in the inner operator, but other solutions could be used, producing similar estimators. Because estimates of this type rely on the approximation or estimation of probability density functions, we refer to them as \emph{pdf-based estimators}.

\cite{PlisBorg14ESTMIM} observe that the kernel-smoothing methods commonly involved in the calculation of pdf-based estimators may induce bias, even at high sample sizes, for simulators with a sparse output. Therefore, they introduce alternative \emph{cdf-based estimators} which rely on the properties of empirical cumulative distribution functions.

Scheff\'e's theorem allows one to write the  $L^1$-distance between two probability density functions in terms of the associated probability functions, as $\int_{\mathcal{Y}}|f_1(y)-f_2(y)|\text{d}y=2(\mathbb P_1(Y\in B)-\mathbb{P}_2(Y\in B))$, where $B$ is the set of values for which $f_1(y)>f_2(y)$. Since $B$ can be written as a union of intervals $(a(t),b(t))_{t=1}^T$, these probabilities can be calculated from the corresponding cumulative distribution functions. Thus, a cdf-based estimator of $\delta_i$ can be obtained as:
\begin{equation}\label{eq:deltafromcdf}
	\widehat \delta_i^\diamond=  \sum_{m=1}^M \frac{n^i_m}{n}\,\sum_{t=1}^{T^i_m}  
	\left(\hat{F}^i_m(\hat{b}^i_m(t)) 
	-\hat{F}^i_m(\hat{a}^i_m(t)) \right)- \left(\hat{F}_Y(\hat{b}^i_m(t)) 
	-\hat{F}_Y(\hat{a}^i_m(t)) \right).
\end{equation}
For further details on the estimation of the intervals $(\hat{a}^i_m(t),\hat{b}^i_m(t)) $, we refer to \cite{PlisBorg14ESTMIM}.

Since $\beta_i$ is itself a cdf-based sensitivity measure, the definition of a one-sample cdf-based estimator is straightforward:
\begin{equation}\label{eq:betaest}
	\widehat{\beta}_{i}^{\diamond}=\sum_{m=1}^{M}\frac{n^i_m}{n}\max_{j\in\{1,\ldots,n\}} \left\vert \hat F_Y(y_j)-\hat F^i_m(y_j) \right\vert,
\end{equation}
where  $\hat F_Y$, and $\hat F^i_m$ are the empirical cdf's of $\mathbf{y} $ and $\mathbf{y}^i_m$, respectively, i.e.:
\begin{equation}\label{eq:emp_cdf}
	\hat{F}_{Y}(y)=\frac{1}{n}\sum_{j=1}^{n}\mathds{1}_{(-\infty,y_j]}(y); \qquad \hat F^i_m(y)=\frac{1}{n^i_m}\sum_{y_j \in \mathbf{y}^i_m }\mathds{1}_{(-\infty,y_j]}(y),
\end{equation}%
and $\mathds{1}_A(y)$ denotes the indicator function, taking the value $1$ if $y\in A$ and $0$ otherwise.

Recalling that the expected value of a random variable $Y$ can be calculated as the integral of its survival function, $\mathbb{E}[Y]=\int_\mathcal{Y}(1-F_Y(y))\text{d}y$, a cdf-based one-sample estimator of the variance-based sensitivity measure, $\eta_{i}$ is given by:
\begin{equation}\label{eq:etafromcdf}
	\widehat{\eta}_{i}^{\Diamond} = \sum_{m=1}^{M}\frac{n^i_{m}}{n}\,\frac{\left( \int_\mathcal{Y} \hat F^i_m(y)-
		\hat{F}_{Y}(y) \text{d}y \right)^2 }{\widehat\sigma^2_Y}.
\end{equation}
Notice that, since the empirical distribution functions are piece-wise constant, the integral in the above expression reduces to a sum. \cite{PlisBorg14ESTMIM} propose an efficient way to calculate this integral.

Most of the estimators found in the literature, including those mentioned above, are constructed either as deterministic approximations or as (frequentist) point estimators. Therefore, quantification of the estimation error (or interval estimation) requires additional manipulation. Finding asymptotic distributions of the estimators in order to provide approximate confidence intervals is not straight forward, except for the variance-based estimator $\eta_{i}^\star$, and even in this case, they are accurate only for high sample sizes. For instance, \cite{GambJano16,janon2014asymptotic,TissPrie15} show that variance-based estimators, calculated with a pick-and-freeze design or a replicated Latin hypercube, are asymptotically normal, but similar results are not available for other sensitivity measures. An alternative for non-deterministic sampling methods, is to replicate the estimation procedure in order to obtain a sample of estimates and corresponding sample-based confidence intervals. This, however requires a number $C>n$ of simulator evaluations and, as mentioned in the introduction, the computational cost of such an effort could be prohibitive for time-demanding realistic applications. The idea of replicates also excludes the use of quasi-random generators to create the probabilistic sensitivity analysis samples, as they are deterministic in nature. As a further alternative, bootstrap confidence intervals have been proposed in the literature in order to avoid the need for additional simulator runs \citep{PlisBorg13,JanoNodePrie14}, but these are not an integral part of the estimation process. 

A second issue to consider when using partition-based one-sample methods is the sample size bias induced by the partition. Quantities related to the marginal distribution of $Y$ are estimated using the full sample size $n$, but those related to the within bin distribution of $Y|X^i\in \mathcal{X}_{m}^i$ are estimated using a smaller sample size $ n^i_m \approx n/M$. While a sample size correction is implicit in the estimation of variances (see Eq. \ref{eq_etaOS}), the same is not true for the pdf and cdf estimates of equations \eqref{eq_deltaPDF} to \eqref{eq:deltafromcdf}. In other words, there is a different granularity when estimating the conditional and the unconditional distributions. In Section \ref{sec:BNPCDF}, we propose two partition-dependent Bayesian estimators which mitigate the sample size bias, while providing a natural way to quantify the estimation error, allowing interval estimation.

Within the Bayesian paradigm, unknown objects are treated as random, and assigned a prior probability measure which reflects the analyst's uncertainty about their values. In this context, \cite{OaklOhag04} treat the input-output mapping $g$ as unknown (at least before evaluation). Thus, they define a semi-parametric regression model with a Gaussian process prior, which allows posterior inference on variance-based sensitivity measures. In fact, it is possible to calculate posterior means for the conditional and unconditional variance of $Y$ and $Y|X^i$ respectively, either analytically or via numerical integration. The approach eliminates the need for a partition of the covariate space, thus solving the second issue mentioned above. However, the posterior distributions of the variance-based measures (e.g. $\eta_{i}$) are not available analytically, and finding a posteriori credibility intervals for estimation error quantification would be cumbersome and this aspect is not treated in the paper. Furthermore, it is not clear how to extend the results to the estimation of other (pdf or cdf-based) sensitivity measures. In Section \ref{sec:BNPPDF}, we present two alternative partition-free Bayesian models which allow interval estimation for these types of sensitivity measures as well. For illustrative purposes, we focus on estimation of the three measures in Table \ref{table_SenMeas}.

\section{Bayesian non-parametric partition-dependent estimation}\label{sec:BNPCDF}

We propose to quantify the uncertainty about fixed but unknown sensitivity measures, $\xi_i$, before (a priori) and after (a posteriori) the observation of a sample, $\{(\mathbf{x}_j,y_j):j=1,\ldots n\}$, within the Bayesian paradigm. The $\xi_i$ play the role of parameters of interest and they are linked to the data through functionals of the marginal and conditional distributions of $Y$ and $Y|X^i$. In view of this, it seems sensible to induce a prior on $\xi_i$ by assigning a prior to the family $\mathcal{P}_i=\{\mathbb{P}_{Y|X^i=x^i}:x^i\in\mathcal{X}^i\}$ of conditional probability measures. Notice that, since $\mathbb{P}_{X^i}$ is assumed known, the marginal distribution of $Y$,  $\mathbb{P}_Y(y) = \int_{\mathcal{X}^i} \mathbb{P}_{Y|X^i=x^i} (y|x^i) \text{d} \mathbb{P}_{X^i} (x^i)$, is fully determined by $\mathbb{P}_{Y|X^i}$, so no additional prior specification is required. For each $i$, $\mathcal{P}_i$ is a family of probability measures on $\mathcal{Y}$, indexed by $x^i\in \mathcal{X}^i$, so defining a prior probability on this space is, in principle, not a simple task. Furthermore, the relation between $\xi_{i}$ and $\mathcal{P}_i$ is complex, making it difficult to conceive an adequate parametric prior. In other words, choosing a family of distributions characterized by a finite-dimensional parameter $\theta$, to express an expert's uncertainty about $\xi_{i}$ through some prior on $\theta$ would seem overly restrictive, if not unreasonable. It is known that an inadequate prior may lead to troublesome posterior \citep{freedman1965asymptotic} and hinder the properties of the proposed estimators. A natural alternative is to use a Bayesian non-parametric prior in order to ensure enough flexibility to capture complex data structures. Bayesian non-parametric methods are not restricted to a finite number of parameters to represent a distribution. Generally speaking, they rely on measure-valued stochastic processes to define priors on the space of probability measures of interest. The supports of such priors are wide, ideally covering the full range of all possible distributions, in our case, over $\mathcal{Y}$ \citep[see e.g.][for an extensive discussion on bayesian non-parametric priors, their properties and their use]{Hea10}. 

Our first proposal can be interpreted as a Bayesian refinement of the cdf-based estimators introduced in the previous section and, as such, relies on a partition of the input space. We assume that the distribution of $Y|X^i=x^i$ is identical for every  $x^i\in \mathcal{X}^i_m$, and denote it by $\mathbb{P}^i_m$. In practice, it is enough to assume that $\mathbb{P}_{Y|X^i=x_i}$ can be well approximated in this way. Prior uncertainty is expressed through a prior on the collection $\{\mathbb{P}^i_m\}_{m=1}^M$. For simplicity, we assume that such distributions are independent and identically distributed (i.i.d.), so the problem becomes that of finding a prior which assigns probability $1$ to a large enough set of probability distributions supported on $\mathcal{Y}$. We focus our attention on the \textit{Dirichlet Process} (DP), first introduced by \cite{Fer73} and widely studied in the BNP literature \citep[see e.g.][Chapter 2, for a discussion on its properties]{Hea10}. 
We therefore define, for each $i=1,\ldots k$ the following Bayesian non-parametric model:
\begin{equation}
	Y|(\mathbb{P}^i_m, X^i \in\mathcal{X}^i_m) \sim \mathbb{P}^i_m;\qquad
	\mathbb{P}^i_m \iid \mathcal{DP}(\alpha G),	
\end{equation}
where $\mathcal{DP}(\alpha G)$ denotes a Dirichlet process with base measure $G$ and concentration parameter $\alpha$. The Dirichlet process could be replaced by a more general stick-breaking process, achieving greater flexibility at a similar computational cost \citep[see e.g.][]{IJ01,PY1997,LMP07b}. In this case, the algorithms and proposed estimators would maintain a similar structure so we focus on the Dirichlet process, without loss of generality, in order to use a notation more familiar to a wider audience. With regards to the hypothesis of independence between the $\mathbb{P}^i_m$, it could be removed through the application of recent developments in BNP methods (see \cite{WooGasArc2011a,Tea06,TehJor2010a, CLP2017} and \cite{CamLiOrPr2017}). This, however, would lead to a complication of the estimation algorithms which goes beyond the scope of this paper.

Note that this Bayesian model is coherent, in the sense that it induces a
unique prior over the unconditional distribution of $Y$, whenever the partitions are equiprobable, that is when $\mathbb{P} (X^i\in \mathcal{X}^i_m)=\frac{1}{M}$ for all $i=1,2,...,k$ and $m=1,2,...,M$. In fact, 
\[
\mathbb{P}_{Y}(\cdot |\mathbb{P}_{1:M}^{i})=\sum\limits_{m=1}^{M}\mathbb{P}%
_{m}^{i}(\cdot )\mathbb{P}(X^i\in \mathcal{X}^i_m)=\frac{1}{M}%
\sum\limits_{m=1}^{M}\mathbb{P}_{m}^{i}(\cdot )\text{.}
\]%
Then, by marginalizing, we obtain%
\[
\mathbb{P}_{Y}(\cdot |\alpha G)=\frac{1}{M}\sum\limits_{m=1}^{M}\int 
\mathbb{P}_{m}^{i}(\cdot )\text{d}\mathcal{DP}(\mathbb{P}_{m}^{i}|\alpha G)=\int 
\mathbb{P}(\cdot )\text{d}\mathcal{DP}(\mathbb{P}|\alpha G),
\]%
because $\int \mathbb{P}_{m}^{i}(\cdot )$d$\mathcal{DP}(\mathbb{P}_{m}^{i}|\alpha
G)$ does not depend on $i$ or $m$. In other words, a priori, $\mathbb{P}_{Y}%
\overset{}{\sim }\mathcal{DP}(\alpha G)$, so that the prior for the marginal simulator distribution is also a Dirichlet process. This statement alone, however, provides no information on the probabilistic dependence of $Y$ on $X^i$. Thus, it is not meaningful, by itself, for a sensitivity analysis. 

The posterior for this model, given the simulator input-output realizations ($Data$ for short), $\{(\mathbf{x}_1, y_1),\ldots,(\mathbf{x}_n, y_n)\}$, can be written as follows:  
\begin{equation}\label{eq:posteriorcond}
	Y|(X^i\in \mathcal{X}^i_m,\mathbb{P}_{m}^{i})\sim \mathbb{P}_{m}^{i};\qquad
	\mathbb{P}_{m}^{i}| Data \overset{ind}{\sim }\mathcal{DP}\left((\alpha +n_{m}^{i})\widetilde{G}_{m}^{i}\right),%
\end{equation}%
where 
\begin{equation}\label{eq:PostMeanCond}
	\widetilde{G}_{m}^{i}=\mathbb{E}[\mathbb{P}_{m}^{i}|Data]=\dfrac{\alpha }{%
		\alpha +n_{m}^{i}}G+\dfrac{n_{m}^{i}}{\alpha +n_{m}^{i}}\sum\limits_{y \in 
		\mathbf{y}^{i}_{m}}\frac{1}{n_{m}^{i}}\delta ^{Dirac}(y) \quad.
\end{equation}
Note that the posterior of the marginal for $Y$ can be obtained as: 
\begin{equation}\label{eq:posteriormarg}
	\mathbb{P}_{Y}(\cdot |\alpha G,Data)=\frac{1}{M}\sum\limits_{m=1}^{M}\int 
	\mathbb{P}_{m}^{i}(\cdot )\text{d}\mathcal{DP}(\mathbb{P}_{m}^{i}|(\alpha +n_{m}^{i})%
	\widetilde{G}_{m}^{i}),
\end{equation}
which may depend both on $i$ and $m$. However, the marginal coherence of the model still holds, at least asymptotically. Informally, for an equiprobable partition, $\mathbb{P}(X^i\in \mathcal{X}^i_m)=1/M$, $n_{m}^{i}\simeq n/M$ when the sample size $n$ is sufficiently large, so $\alpha /(\alpha +n_{m}^{i})
\simeq M\alpha/(M\alpha +n)$ and $n_{m}^{i}/(\alpha +n_{m}^{i})\simeq n/(M\alpha +n)$. Furthermore, $\sum
(1/n_{m}^{i})\delta ^{Dirac}(y)\simeq \sum(M/n)\delta ^{Dirac}(y)$. Thus, asymptotically, $\mathbb{P}_{Y}(\cdot
|\alpha G,Data)$ does not depend on $m$ or $i$ and $\mathbb{P}_{Y}(\cdot |\alpha G,Data)\overset{}{\sim }
\mathcal{DP}((\alpha +n)\widetilde{G})$, where 
\begin{equation}\label{eq:postmeanmarg}
	\widetilde{G}=\dfrac{\alpha }{\alpha +n}G+\dfrac{n}{\alpha +n}
	\widehat{\mathbb{P}}_n,
\end{equation}
and $\widehat{\mathbb{P}}_n$ denotes the empirical distribution of $Y$ based on the full set of observations, $(y_1,\ldots,y_n)$. Note that this is the usual posterior corresponding to the DP prior on $\mathbb{P}_Y$.

The sensitivity measures we aim to estimate are functionals of the conditional and marginal distributions. The posterior means in eqs. \eqref{eq:PostMeanCond} and \eqref{eq:postmeanmarg}, respectively, may be proposed as Bayesian estimators of such densities. Thus, a Bayesian point estimator of $ \xi_i $ may be given by:
\begin{equation*}
	\widetilde{\xi }_{i}=\sum_{m=1}^{M} \dfrac{n_m^i}{n} \zeta(\widetilde{G},
	\widetilde{G}_{m}^{i})
\end{equation*}
Unfortunately, the direct calculation of  $ \widetilde{\xi }_{i} $ is impractical. Moreover, our purpose is to provide interval estimation, so as to quantify the uncertainty associated to point estimates. A way out is to sample observations (i.e., predicted realizations of the output) from  $\widetilde{G}$ and $\widetilde{G}_{m}^{i}$, in order to enrich the sample. More specifically, we have a vector $\mathbf{y}$ of $ n $ observations from the original simulator used to estimate $\mathbb{P}_Y$, but only $ n_m^i $ of these belong to $\mathbf{y}^{i}_{m}$ and are therefore used to estimate $\mathbb{P}_m^i$. Because $ n_m^i <n$, the precision issue discussed in Section \ref{subsec:onesample} emerges, causing a bias in the empirical estimation of $\xi_i$. By re-sampling from $\widetilde{G}$ and $\widetilde{G}_{m}^{i}$ we can enlarge both vectors, making them of the same size and, potentially, arbitrarily large. Our proposal here is simply to sample $ n-n_m^i $ observations from $\widetilde{G}_{m}^{i}$, thus obtaining two vectors of size $n$. The intuition underlying this corresponds to the non-parametric \textit{Bayesian bootstrap} (Bb)  \citep{hjort1985bayesian,hjort1991bayesian}. 
In our case, for each $m$ a sample $\widetilde{\mathbf y}_m^i=\{\widetilde{y}_{n_m^i+1}^i,\dots,\widetilde{y}_{n}^i\}$ of size $n-n_m^i$ is obtained from $\widetilde{G}_{m}^{i}$.
A value of $\widehat{\xi}^{Bb,s}_{i}$ in Eq. \eqref{eq_xiOS} can be calculated through any of the methods discussed in Section \ref{subsec:onesample}, using $\mathbf y$ to estimate all quantities related to the marginal distribution of $Y$ and the extended vector $\mathbf{y}_m^{Bb,i,s} = (\mathbf y_m^i, \widetilde{\mathbf y}_m^{i,s})$ to estimate all quantities related to the conditional distribution of $Y|X^i\in\mathcal X_m^i$. Informally, the weighted average over $ m $ can be seen as approximately simulated from the posterior distribution of $\xi_{i}$. By repeating this procedure $ S $ times, we obtain a Bb sample  $\{\widehat{\xi }_{i}^{Bb,s}: s=1,2,...,S\}$. We propose the Monte Carlo average:
\begin{equation*}\label{eq:Bbestimator}
	\widehat{\xi}_i^{Bb} = \frac{1}{S}\sum_{s=1}^{S}\widehat{\xi }_{i}^{Bb,s}
\end{equation*}
as a point estimator of $\xi_i$. Approximate credibility intervals can be obtained from the empirical quantiles. 
Note that, because each $\widetilde{y}_j^i$ is simulated from a single distribution, $\widetilde{G}_{m}^{i}$, the sampling process can be done in parallel and the method is computationally fast. However, the uncertainty is underestimated because the additional variability captured by the posterior distribution of Eq. \eqref{eq:posteriorcond} is ignored. 

A more accurate alternative is to sample $\widetilde{\mathbf y}_m^i$ jointly from the Dirichlet process posterior distribution \eqref{eq:posteriorcond}, instead of sampling each $\widetilde{y}_j^i$ from the posterior mean. This can be done via the \textit{P\'olya Urn scheme} (Pu) of \cite{blackwell1973ferguson}. Specifically, $\widetilde{\mathbf y}_m^i$ is generated as a realization of the P\'olya sequence:  
\begin{equation}
	\widetilde{Y}^i_{j+1} | \left(\widetilde{y}^i_{n_m^i+1:j}, Data\right) \sim \frac{\alpha}{\alpha+j} G + \frac{j}{\alpha+j} \widehat{\mathbb{P}}_{j} \quad \forall\, j\ge n_m^i.
	\label{eq_polya}
\end{equation}
Once again, the extended samples $\mathbf{y}_m^{Pu,i,s} = (\mathbf y_m^i, \widetilde{\mathbf y}_m^{i,s})$ can be used to obtain a value $\widehat{\xi}^{Pu,s}_{i}$ by any available method to calculate the expression in Eq. \eqref{eq_xiOS}. We use $
\widehat{\xi}^{Pu}_i$ to denote the Monte Carlo average of a sample of size $S$ generated in this way. Note that this is a point estimator with the same expectation as $\widehat{\xi}^{Bb}_i$. However, a greater variability which fully accounts for the uncertainty on $\mathbb{P}_m^i$ results in wider credibility intervals. The sampling procedure is now sequential for $s=1,\ldots,S$, so the price for greater accuracy in uncertainty estimation is a slightly higher computational time.  

The technical details for Bb and Pu estimators are presented in Section \ref{alg_OSBB} (Appendix \ref{subsec:algos}).

\subsection{Simulation study}\label{subsec:Exampls}
We illustrate the performance of the two classes of estimators proposed above, via two toy examples for which the sensitivity measures can be calculated analytically (see Table \ref{tab_AnalyGSvalues} columns $1$ and $2$ ). The first example is the 2-input simulator 
\begin{equation}
	Y=\frac{X^1}{X^1+X^2},\label{eq:NLNA}
\end{equation}
where $X^1$,$X^2 \overset{iid}{\sim} \text{Gamma}(3,1)$, so that the output $Y$ follows a Beta distribution.   The second example is the 21-input simulator
\begin{equation}
	Y=\sum_{i=1}^{21} a_i X^i,  \label{eq:GauMod}
\end{equation}
where $X^i \sim \text{Normal} (1,1)$, with $a_1 = \dots = a_7= -4$, $a_8 = \dots = a_{14} = 2$, and $a_{15} = \dots = a_{21} = 1$. The 21 inputs are correlated with $\rho_{i,j} = 0.5$. Therefore, Model inputs with indices in $1,2 \dots 7$
are the most important, followed by inputs with indices in $15,16,\dots,21$, followed by inputs with
indices in $8,9\dots 14$. Columns $3$ to $6$ in Table \ref{tab_AnalyGSvalues} displays the analytical values of the sensitivity measures.

\begin{table}[h]
	\centering
	\caption{Analytical values of $ \eta_i $, $ \delta_i $ and $ \beta_i $  for the two test simulators used in this section}
	\label{tab_AnalyGSvalues}      
	\begin{tabular}{c |cc |ccc}
		\hline\noalign{\smallskip}
		& \multicolumn{2}{c}{2-input}& \multicolumn{3}{|c}{21-input} \\ 
		\hline
		Measure & $X_1$ &$X_2$ & $X_1-X_7$ & $X_8-X_{14}$ & $X_{15}-X_{21}$ \\ 
		\hline
		$\eta_i$ & 0.496 & 0.496 & 0.309 & 0.064 & 0.092\\
		$\delta_i$  & 0.315 & 0.315 & 0.212 & 0.084 & 0.102\\ 
		$\beta_i$& 0.289& 0.289& 0.205 & 0.083 & 0.101\\
		\noalign{\smallskip}\hline
	\end{tabular}
\end{table}

\begin{figure*}
	\centering
	\includegraphics[width=0.75\textwidth]{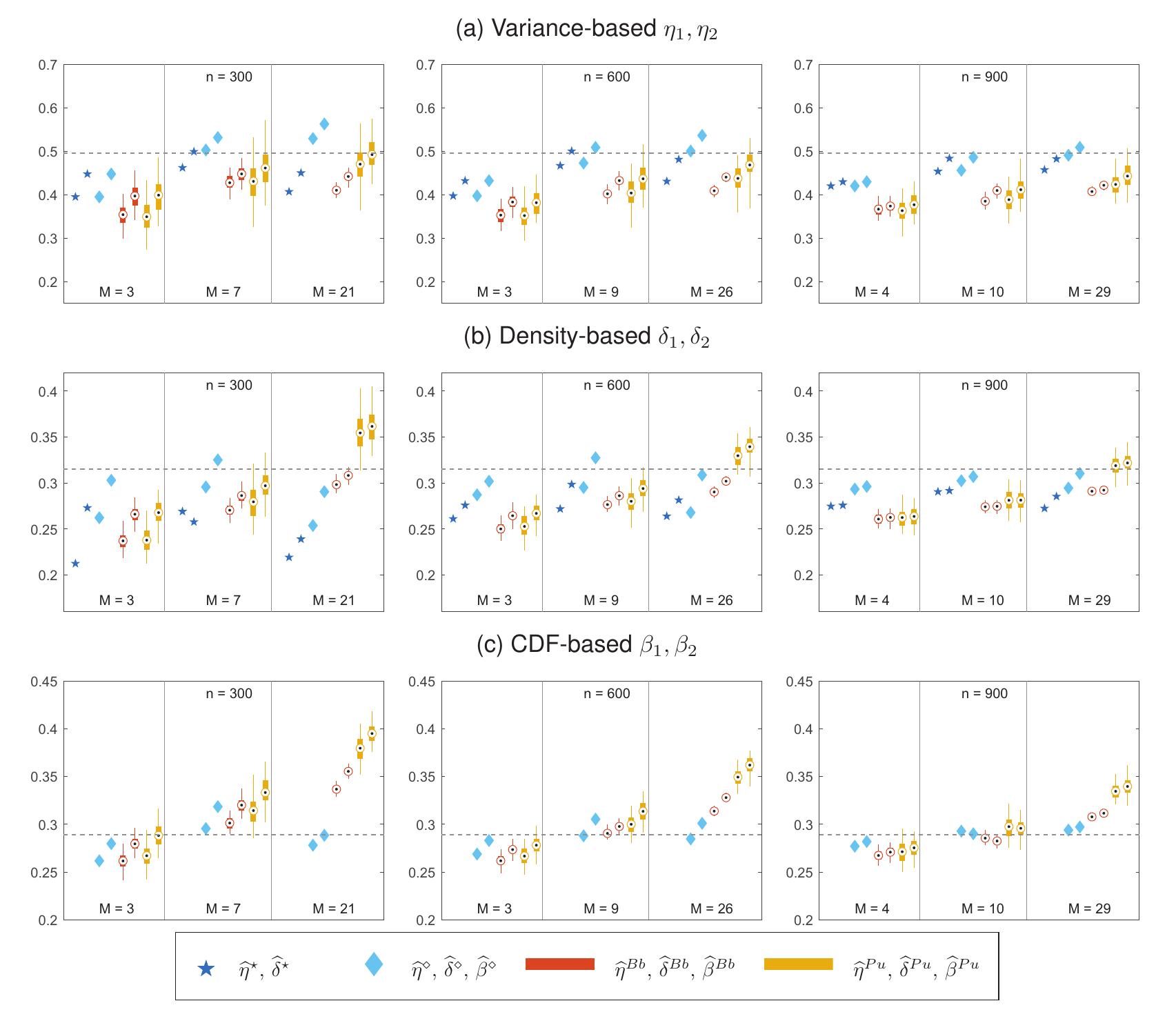}
	\caption{Results for the 2-input simulator in Eq. \eqref{eq:NLNA}: Comparison of sensitivity measures estimates using frequintist  pdf/cdf-based estimators and partition-dependent Bayesian non-parametric estimators.Bayesian estimates include 95\% credibility intervals.}
	\label{fig:NLNA}        
\end{figure*}

We are interested in small sample sizes, which make the estimation of global sensitivity measures challenging. In particular, we consider $ n=\{300, 600, 900 \}$.  The input data, $\mathbf{x}$, is generated via Quasi-Monte Carlo. For each $ n $, alternative choices of the partition size, $M$, are explored. The mass parameter, $ \alpha $, for the DP prior is set equal to $ 0.1 n/M$ throughout. The base measure, $ G $, is chosen in correspondence with the support of $ Y $: a Beta distribution for the first example and a Normal distribution for the second; the hyper-parameters are fixed through an empirical approach, based on the available sample $\mathbf{y}$. Note that this choice centres the prior distribution for $Y|X^i\in\mathcal{X}_m^i$ roughly around the marginal distribution of $Y$, thus favouring, a priori, independence between the $Y$ and $X^i$, with a precision proportional to the number of observations in each partition set. In practical applications, prior information elicited from experts may be expressed through different choices of $\alpha$ and $G$.

\begin{figure*}
	\centering
	\includegraphics[width=0.75\textwidth]{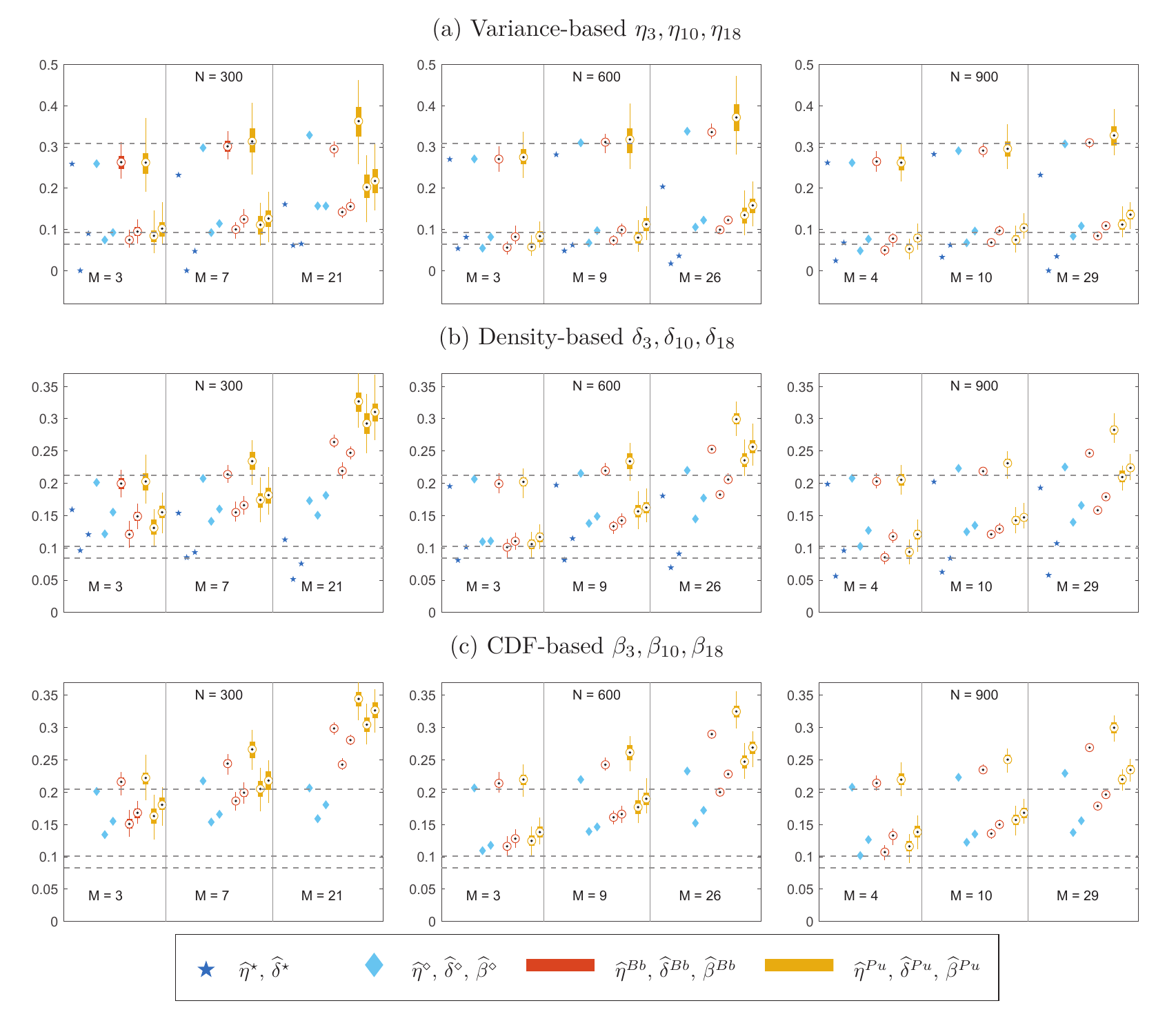}
	\caption{Results for the $21-$input simulator in Eq. \eqref{eq:GauMod}: Comparison of sensitivity measures estimates using frequintist pdf/cdf-based estimators and partition-dependent Bayesian non-parametric estimators. Bayesian estimates include 95\% credibility intervals.}
	\label{fig:GauMod2}        
\end{figure*}

We compare the Bayesian bootstrap and P\'olya urn estimators to traditional point estimators for three global sensitivity measures. Results are reported in Fig.s \ref{fig:NLNA} and \ref{fig:GauMod2}: the first row corresponds to $ \eta_i $, the second to $ \delta_i $ and the third to $\beta_i$.  Columns, from left to right, correspond to increasing sample sizes. Each graph is divided into three blocks displaying importance measures estimates based on alternative choices of $M$. The dotted lines display the analytical values.

We first consider the left-most panel of Fig. \ref{fig:NLNA}(a). At $ n=300 $ the estimates vary notably with the partition size: they are downward biased for $ M=3 $ and upward biased for $ M=21 $. Observe that at $ M=21 $, we have $n_m^i\simeq 9$, a number too small to be reasonably chosen by the analyst.  However, the bias is systematic, that is, it affects identically all estimates. Estimates are less affected by the partition choice as the sample size increases. Recall that in realistic applications, where an analyst would not know the true values of the sensitivity measures, the main interest is on the ordinal ranking of the inputs. In this example, $X^1$ and $X^2$ are equally important. However, looking at the point estimators $ \widehat{\eta}_i^{\star} $ and $ \widehat{\eta}_i^\diamond $ the analyst would rank $ X^2 $ as more important than $ X^1 $ for most combinations of $n$ and $M$. The credibility intervals for $ \widehat{\eta}_i^{\text{Pu}} $ display a large overlapping that would prevent the analyst from ranking $ X^2 $ above $ X^1 $: there is too much uncertainty in the estimates to make such conclusion. Notice the underestimation of the uncertainty surrounding $ \widehat{\eta}_i^{\text{Bb}} $. Rows (b) and (c) of Fig. \ref{fig:NLNA} show a similar behaviour for the $\delta_i $ and $ \beta_i $ sensitivity measures.\\

Fig. \ref{fig:GauMod2} shows the estimates for the $21-$input simulator in \eqref{eq:GauMod}. For a better display clarity, instead of reporting seven sensitivity measures per group, we show numerical values for a representative of each input group, namely $X^3, X^{10}$ and $X^{18}$. 
For variance-based sensitivity measures (Fig. \ref{fig:GauMod2}(a) ) for all $n$ and $M$ considered, $ \widehat{\eta}_i^{\star} $, $ \widehat{\eta}_i^\diamond $, $ \widehat{\eta}_i^{\text{Bb}} $ and $ \widehat{\eta}_i^{\text{Pu}} $ are able to correctly identify $  X^3 $ as the most relevant variable. Regarding $ \delta_i $ and $ \beta_i$ (Fig. \ref{fig:GauMod2}(b) and Fig. \ref{fig:GauMod2}(c), respectively), one identifies $ X_3 $ as the most important input in almost all combinations of sample sizes and partition selections. The ranking becomes unclear for $ n=300 $ and $ M=21 $. However, this choice would leave  about $9$ realizations per partition, a number too small to be reasonably chosen by the analyst. For the remaining group of inputs, the overlapping error bands for the $Bb$ and $Pu$ estimates would not allow us to deem  $ X^{10} $ more relevant than $X^{18}$, with either $ \eta_i $,$ \delta_i $ or $ \beta_i $. Thus higher sample sizes would be needed for neatly ranking the second and third most important groups of simulator inputs.

Overall, Fig.s \ref{fig:NLNA} and \ref{fig:GauMod2} suggest that the proposed estimators allow the identification of the most important inputs, even at small sample sizes, and, most relevantly, they provide a measure of the uncertainty in the assessment. However, the results also display a strong dependence on the partition size $M$. While \textit{i)} as observed in \cite{StroOakl13} (see their Fig. 1, at p. 759), the importance of selecting an optimal partition size diminishes as the sample size increases and \textit{ii)} a suboptimal partition selection has in most cases an identical impact on the sensitivity measures (i.e., the sensitivity measures of all inputs are simultaneously upward or downward biased), the analyst is still left with the question of what is the optimal partition size for a given sample. Unfortunately, there seems to be no universally optimal selection rule (see Appendix \ref{Sec:AppHeuristic}). Clearly, the problem would be solved if partition-independent estimators were available. In the next section, we study two proposals of Bayesian estimators that avoid the partition choice problem.

\section{Bayesian non-parametric partition-free estimation}\label{sec:BNPPDF}
In this section, we propose two classes of Bayesian partition-free estimators. The first is based on the use of an infinite mixture model to estimate the joint density of $Y$ and $X^i$. The second, uses a Bayesian non-parametric regression model to estimate the conditional density of $ Y $ given $ X^i $. 

\subsection{Joint density-based estimation \label{sec_DGM}} 
The intuition is that all sensitivity measures under consideration can be recovered from the joint distribution of $Y$ and $X^i$. Therefore, in order to do Bayesian inference on $ \xi_i $ it suffices to place a prior on the joint density $ f_{X^i,Y}$. We propose to do so by means of a nonparametric mixture model (see, e.g. \citet{ferguson1983bayesian, lo1984class}). In other words, we consider $f_{X^i,Y}$ to be defined as a mixture:
\begin{equation}
	f_{X^i,Y}(\cdot,\cdot)|P = \int \mathcal{K}(\cdot,\cdot\vert \theta) \text{d}P(\theta),
	\label{eq_fX}
\end{equation}
where $\mathcal{K}$ is a parametric bivariate density and the mixing measure $ P$ is a probability distribution over an appropriate space of parameters. The model is completed by assigning a non-parametric prior, $\Pi$, on $P$.  Most common choices of $\Pi$ assign probability one to discrete distributions of the form
\begin{equation}\label{eq:jointwkp}
	P(\theta)= \sum_{\ell=1}^{\infty}w_\ell\, \delta^{Dirac}(\theta_\ell),
\end{equation}
placing mass $ w_\ell $ on locations $ (\theta_\ell) $. In the literature, particular attention has been paid to nonparametric priors admitting a stick-breaking construction \citep{Pitman96, sethuraman1994constructive} where the weights $ \underline{w}=(w_1,w_2,...)$ are defined as realization of random variables satisfying
\begin{equation}\label{eq:stick-breaking}
	W_1 = V_1, \quad W_\ell = V_\ell\prod_{\ell' =1}^{\ell -1} (1-V_{\ell'})
\end{equation}
and independent of $\underline{\theta}=(\theta_1,\theta_2,...)\iid G $. Rich families of stick-breaking priors can be defined via different distributional assignments for the sequence $(V_1, V_2, \ldots)$ \citep[see, e.g.,]{FLP2012, IJ01}. The main advantage over other types of construction is that the stick-breaking representation of the random weights allows for efficient simulation algorithms, specially in the context of nonparametric mixture models \citep{IJ01, PapRoberts08, KGW11, Yauetal2011}. However, the most popular stickbreaking prior remains the Dirichlet process, well known even outside the specialized community of Bayesian nonparametrics. For this reason, we will focus our analysis on DP mixtures, thus letting $P\sim \Pi =\mathcal{DP}(\alpha G)$. Additionally, for simplicity, we choose $ \mathcal{K}$ to be a bivariate normal density, following the density estimation scheme of \cite{escobar1995bayesian}.
In this case, $\theta_\ell=(\mu_\ell,\Sigma_\ell)$ and, to simplify calculations, we select $G$ as a conjugate Normal inverse-Wishart distribution. Thus, the the integral in \eqref{eq_fX} reduces to a sum and  the joint density can be written as:
\begin{equation}\label{eq:jointwk}
	f_{X^i,Y}(\cdot,\cdot)\vert P= \sum_{\ell=1}^{\infty}w_\ell \cdot \mathcal{N}
	(\cdot,\cdot|\mu_\ell,\Sigma_\ell),
\end{equation}
where the weights follow \eqref{eq:stick-breaking}, with $V_i \iid \text{Beta}(1,\alpha)$.

Inference on this model is usually achieved via an MCMC scheme resulting in a sample from the posterior distribution of $f_{X^i,Y}$ given the $Data$. In the case of the DP-mixture, the function \texttt{DPdensity} from the \texttt{R} package \texttt{DPpackage} provides an off-the-rack solution. In practice, the MCMC scheme generates, at each iteration $s=1,\ldots,S$, values $(\underline{w}^s,\underline{\mu}^s,\underline{\Sigma}^s)$ which, substituted in expression \eqref{eq:jointwk}, produce a density function, $f^{BNJ,s}_{X^i,Y}$. Analytical expressions for the marginal and conditional densities,  $f^{BNJ,s}_Y$ and $f^{BNJ,s}_{Y|X^i}$ as mixtures of normal distributions are made easily available by the choice of the Gaussian kernel. Clearly, it is also possible to evaluate the corresponding cumulative distribution functions. Thus, it is possible compute the global sensitivity measures of interest, $ \eta_i^{BNJ,s} $, $ \delta_i^{BNJ,s} $, $ \beta_i^{BNJ,s} $ from their definitions (Table \ref{table_SenMeas}),  obtaining a posterior sample of each. We denote the sample means by $\widehat{\eta}^{BNJ}_i$, $ \widehat{\delta}^{BNJ}_i $ and $ \widehat{\beta}^{BNJ}_i $, respectively, proposing them as Bayesian point estimators. Approximate credibility intervals can be obtained from the empirical quantiles of the samples. The procedure is summarized in Section \ref{alg_BNJ} of Appendix \ref{subsec:algos}, to which we refer for further details.

It is important to observe that the known marginal distribution for $X$ does not, in general, coincide with the marginal distribution for $X$ derived from each $f^{BNJ,s}_{X^i,Y}$. Thus, by using only the joint density $ f_{X^i,Y}$ to estimate the sensitivity measures, important information, standard in global sensitivity analysis is wasted. In fact, inference for conditional densities based in the joint model is known to be approximate \citep[see e.g.][]{MuQ2004}. In the next section, we present an alternative estimation method which avoids this problem through a recent Bayesian approach to conditional density estimation.

\subsection{Conditional density-based estimation}\label{sec:CDB}
We now propose to use a Bayesian non-parametric regression model to do inference directly on the conditional density of $Y|X^i$, thus using all of the information contained in the $Data$ to estimate the relationship between the variables and exploiting the knowledge of the marginal distribution of $X$ to obtain the marginal distribution of $Y$. The idea is to transform the non-parametric mixture of equation \eqref{eq:jointwk} into a mixture of conditional densities:  
\begin{equation}
	f_{Y|X}(y|x) = \int \mathcal{K}(y|x,\theta) \text{d} P_x (\theta),
\end{equation}
This time a non-parametric prior, $\Pi$, is placed on the family, $\{P_x\}_{x\in \mathcal{X}}$ of mixing distributions indexed by $x$. Analogous to the DP mixture model of the previous section, a dependent DP mixture model or DDP mixture \citep{maceachern1999dependent,maceachern2000dependent} is obtained when $P_x$ follows a DP prior, marginally for every $x$, so that:
\begin{equation}
	\mathbb{P}_x (\theta) = \sum_{\ell=1}^{\infty} w_\ell(x) \delta_{\theta_\ell(x)}.
	\label{eq:fyx}
\end{equation}
The random covariate-dependent weights $W_\ell (x)$ follow the stick-breaking construction of Eq. \eqref{eq:stick-breaking}, for i.i.d. random processes $\{V_\ell(x):x\in\mathcal{X}\}$. In other words, $\mathbf V(x) \sim \mathcal{DP}$ for every $x$. It has been proved sufficient flexibility is achieved through models in which only the particles $\theta_\ell$ or the weights $w_\ell$ depend on the covariate $x$ \citep{BJQ12}, the second option being favoured due to better predictive capabilities. Several proposals have been studied in the literature, focusing on alternative definitions of the random functional weights $w_\ell (x)$ \citep[e.g.][]{DuP08, GrS06, RoD11}. 

The stick-breaking structure of the weights, which imposes a geometric decay, may be bypassed through an alternative construction allowing further flexibility:
\begin{equation}\label{eq:weightsIsa}
	w_\ell(x) = \frac{\omega_\ell \mathcal{K}(x|\psi_\ell)}{\sum_{\ell'=1}^{\infty} \omega_{\ell'} \mathcal{K}(x|\psi_{\ell'}) }.
\end{equation}
The denominator of this expression is, again, an infinite mixture of parametric kernels, $\mathcal{K}$, this time with support $\mathcal{X}$. Each $\omega_\ell$ can be interpreted as the probability that a realization of $ Y $ comes from the $\ell$-th regression component regardless of the value of $X$, just as $\omega_\ell$ is the conditional probability given $X = x$. Such density regression model, where the weights $w_\ell$ in \eqref{eq:weightsIsa} follow the stick-breaking representation of \eqref{eq:stick-breaking} and the extended parameters $(\theta_\ell,\psi_{\ell})$ are i.i.d. from some adequate base measure, $G$, was proposed by \cite{antoniano2014bayesian}, to which we refer the reader for additional details on the role and choice of hyper parameters, as well as the algorithm used for inference. 

We adopt this construction to estimate the conditional density $f_{Y|X^i} (y|x_i)$ as a mixture of linear regression models:
\begin{equation}\label{eq:mixcondIsa}
	f_{Y|X^i} (y|x^i) = \sum_{\ell=1}^{\infty} w_\ell(x^i) \mathcal{N}(y|a_\ell+b_\ell  x^i, \sigma_\ell),
\end{equation}
where $w_\ell(x_i)$ is given by Eq. \eqref{eq:weightsIsa}, with a DP prior. Once again, a MCMC approach is used to generate a sample, this time from the posterior distribution of $f_{Y|X^i}$. Each $ f^{BNC,s}_{Y|X^i} (y|x_i), s = 1, \ldots, S$, together with the known marginal for $X^i$ can be used to calculate (e.g. by numerical integration) a corresponding marginal for $ Y $. As discussed in Section \ref{sec_DGM}, this is all that is needed to compute the global sensitivity measures of interest, $ \eta_i^{BNC,s} $, $ \delta_i^{BNC,s} $ and $ \beta_i^{BNC,s} $. These, again allow point estimation, e.g. via the Monte Carlo averages, which we denote by  $ \widehat{\eta}_i^{BNC} $, $ \widehat{\delta}_i^{BNC}  $ and $ \widehat{\beta}_i^{BNC}  $, and interval estimation, via empirical quantiles. Section \ref{alg_BNC} (Appendix \ref{subsec:algos}) summarizes the procedure and offers additional technical details.

\begin{figure}
	\begin{turn}{90}
		\begin{minipage}[c][\textwidth][c]{\textheight}
			\subfigure[Results for input $X^1, X^2$ the $2-$input simulator in Eq. \eqref{eq:NLNA}\label{fig_BNP_analy}]{\includegraphics[width=.5\textwidth]{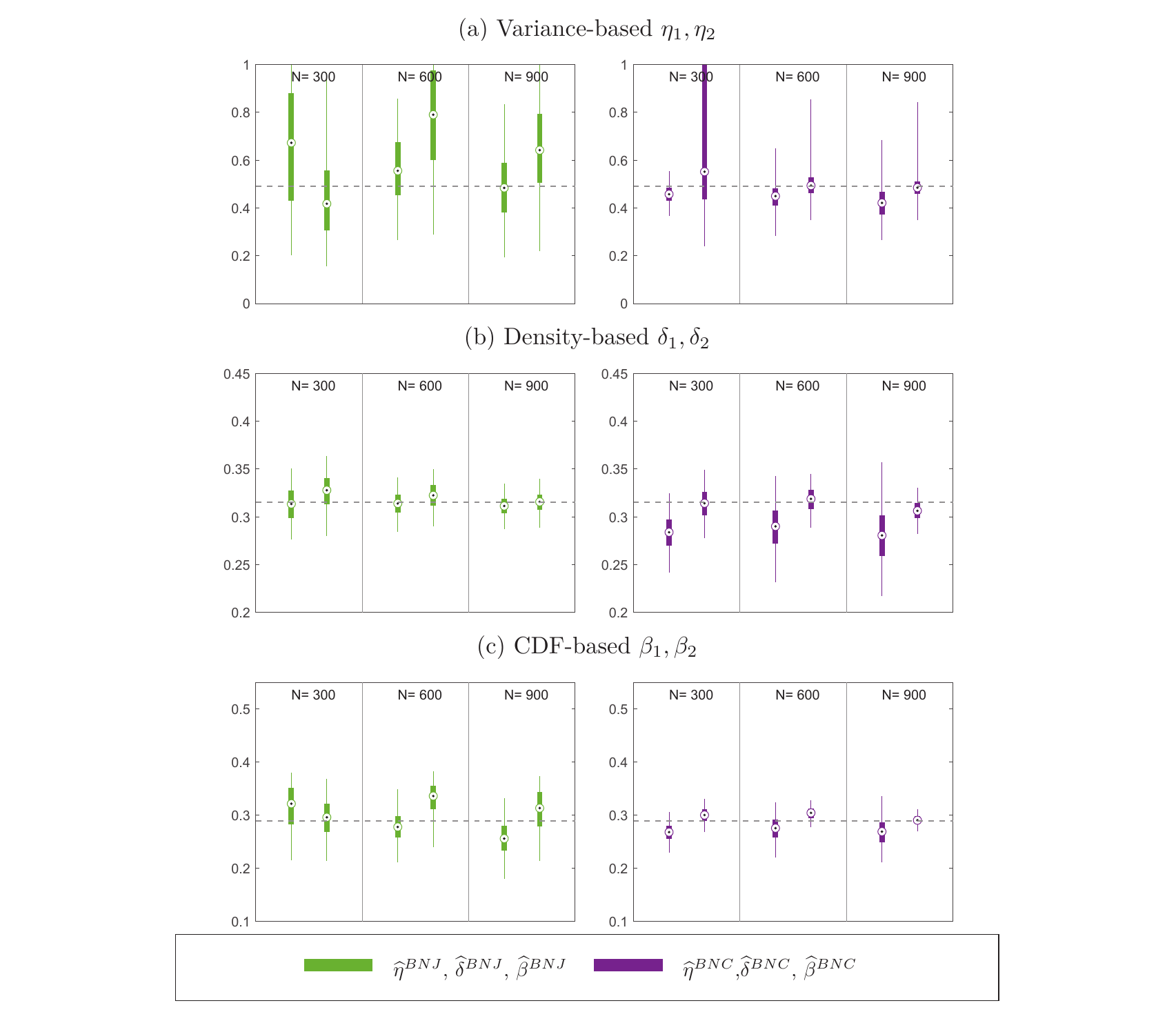}}\hspace{-5em}
			\subfigure[Results for input $X^3,X^{10},X^{17}$ of the $21-$input simulator in Eq. \eqref{eq:GauMod}\label{fig_BNP_AG}]{\includegraphics[width=.5\textwidth]{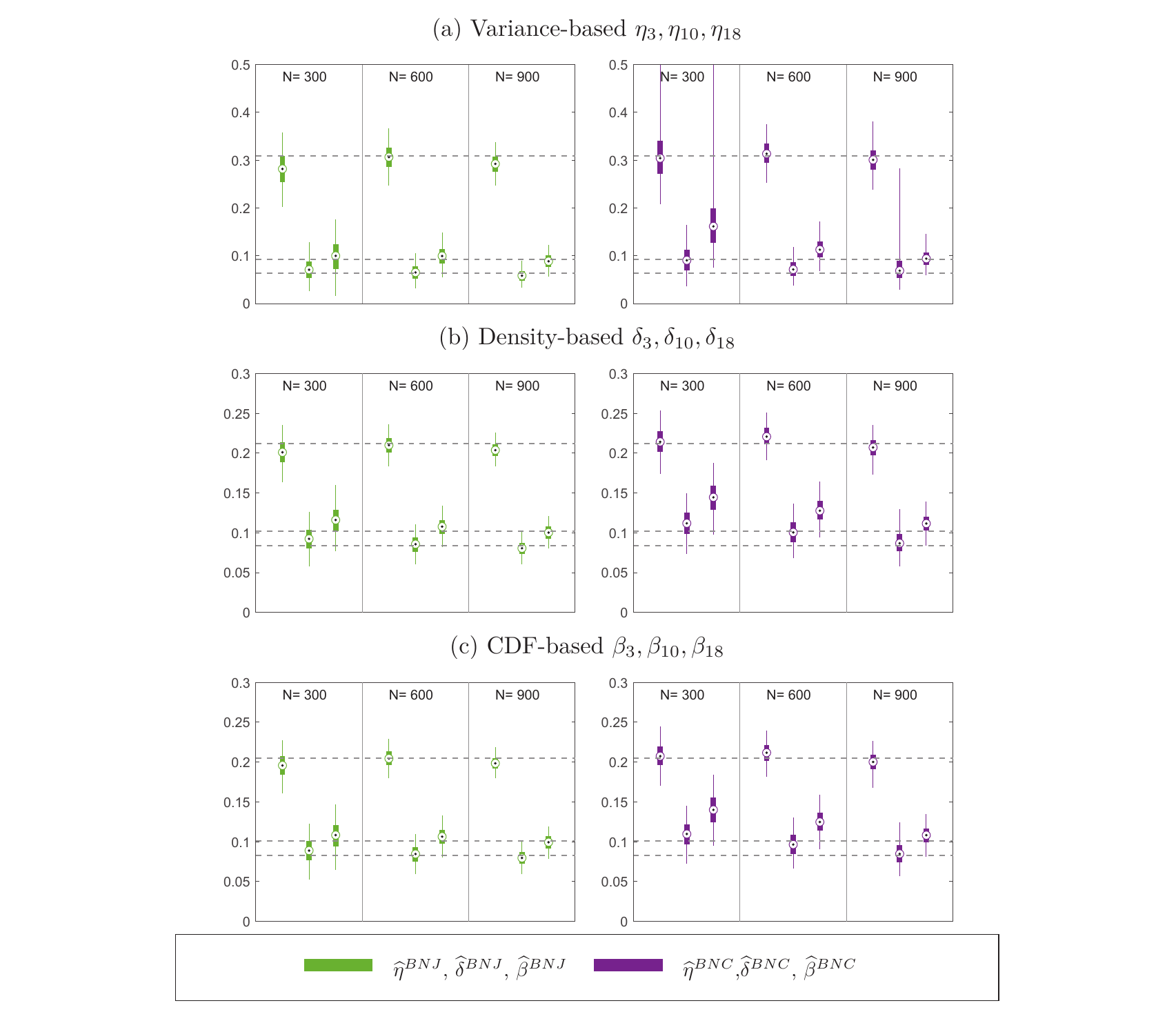}}
			\caption{Comparison of sensitivity measures estimates with 95\% credibility intervals using Bayesian non-parametric partition-free joint/conditional estimators. The dash lines are analytical values of sensitivity measures in Table \ref{tab_AnalyGSvalues}.\label{fig_BNP_toy}}	
		\end{minipage}		
	\end{turn}
\end{figure}
\subsection{Simulation study}\label{sec:simstud}
We examine the performance of the classes of partition-independent estimators proposed in Sections  \ref{sec_DGM} and  \ref{sec:CDB}, via the $ 2- $input and $ 21- $input simulators introduced in Section \ref{subsec:Exampls}.
For both joint and conditional density-based estimation, we set a burn-in period as $10n$ and the stored MCMC samples size $S=1000$. Results are illustrated in Fig. \ref{fig_BNP_toy}. 

Figure \ref{fig_BNP_AG} reports results for the $21-$input simulator.
The Bayesian non-parametric joint estimators $\widehat{\eta}^{BNJ}_i$, $\widehat{\delta}^{BNJ}_i$ and $\widehat{\beta}^{BNJ}_i$ correctly recover the true values of the parameters and, as the sample size increases from $n=300$ to $n=600$, the credibility intervals become narrower. At $n=900$, there is no more overlap among the three groups of sensitivity measures, allowing the analyst to rank the inputs neatly. 
Regarding the Bayesian non-parametric conditional estimates, we observe that $ X_3 $  is correctly identified as the most relevant input at all sample sizes. The values  $ \eta_{10}^{BNC}, \delta_{10}^{BNC} $, $ \beta_{10}^{BNC} $ as well $ \eta_{18}^{BNC}, \delta_{18}^{BNC} $, $ \beta_{18}^{BNC} $ are overestimated by the $BNC$ estimators at $ n=300 $. However, the bias is reduced as $ n $ increases and at $ n=900 $ the credibility intervals appear centred around the analytical values of the sensitivity measures. 
We also observe that for both classes of estimates the analytical value of the sensitivity measures falls within the $ 95\%  $ credible intervals.

For this example joint Bayesian estimators seem to outperform their conditional counterpart. This is to be expected, because the joint Gaussian structure of the data is more easily recovered by the joint model in this case, so the loss due to ignoring the true distribution of $X^i$ has a lesser effect on the results. However, we can appreciate a reassuring improvement of the BNC estimates as the sample size increases. One may argue that, in a situation in which the true conditional distribution of $Y$ given $X^i$ is unknown and may be complex, estimation based on the conditional density model may be preferred, as more robust; the price to pay is that a larger sample size may be required, specially in high-dimensional situations. We then challenge these results for $ 2- $input simulator in Eq. \eqref{eq:NLNA}, in which the distributions are not normal.

Assume for the moment that the analyst does not know the true value of the sensitivity measures. In terms of ordinal ranking, Fig.~\ref{fig_BNP_analy} suggests that the two simulator inputs are equally important.  The credibility intervals of $ X_1 $ and $ X_2 $ obtained with both the BNJ and BNC estimators are overlapping at the all sample sizes and for all sensitivity measures, so that the analyst cannot deem one of them more important than the other. The performance of the two estimators is similar. However, note that the credible intervals of the joint model (BNJ) are wider than those for the conditional model especially for variance-based sensitivity measures. As expected, for a non-normal distribution, the joint model resents from the wrongful introduction of the marginal distribution for $X_i$. We analyze this behavior further in addressing results for the LevelE case study.

\section{Case study: LevelE simulator}\label{sec:levelE}
\begin{table*}
	\centering
	\caption{Simulator inputs for the LevelE code. $ U(\cdot,\cdot) $ and $LU(\cdot,\cdot) $ stand for the uniform and log-uniform distributions respectively}
	\label{tab_LevelEinput}      
	\begin{tabular}{lll}
		\hline\noalign{\smallskip}
		Input & Meaning & Distribution  \\
		\noalign{\smallskip}\hline\noalign{\smallskip}
		$X_1$ & Containment time  & U(100,1000) \\
		$X_2$ & Iodine Leach rate & LU($10^{-3}, 10^{-2}$)\\
		$X_3$ & Neptunium chain Leach rate & LU($10^{-6}, 10^{-5}$)\\
		$X_4$ & Iodine retention factor (1st layer) & LU($10^{-3}, 10^{-1}$)\\
		$X_5$ & Geosphere water velocity 1st layer & U(100,500) \\
		$X_6$ & Geosphere Length 1st layer & U(1,5)\\
		$X_7$ & Factor to compute Neptunium retention coefficients Layer 1 & U(3,30)\\
		$X_8$ & water velocity in geosphere's 2nd layer  & LU($10^{-2}, 10^{-1}$)\\
		$X_9$ & Length of geosphere's 2nd layer & U(50,200)\\
		$X_{10}$ & Retention factor for I (2nd layer) & U(1,5)\\
		$X_{11}$ & Factor to compute Neptunium retention coefficients Layer 2 & U(3,30)\\
		$X_{12}$ & Stream flow rate & LU($10^{5}, 10^{7}$)\\
		\noalign{\smallskip}\hline
	\end{tabular}
\end{table*}

In this section, we evaluate the performance of the proposed estimators through the benchmark simulator of sensitivity analysis, LevelE. The LevelE code simulates the release of radiological dose from a nuclear waste disposal site to humans over geological eras. The code has been developed in an international exercise launched by the Nuclear Energy Agency (NEA) in the mid 1980's \cite{NuclearEnergyAgency1989}. Goal of the exercise was the realization of a reference simulator for the prediction of flow and transport of radionuclides in actual geologic formations against which to compare other simulators developed internationally to support the selection of radioactive wast management policies. Since then, LevelE has become the benchmark simulator of sensitivity analysis  \citep{Saltelli2000, saltelli_jasa_2002}. During the international exercise, distributions for the uncertain simulator inputs were assessed (Table \ref{tab_LevelEinput}), and have become the reference for analysis on this code.
From a technical viewpoint, the LevelE code solves of a set of nested partial differential equations that compute the released radiological dose in Sievert/year over a time range of $ t=10,000 $ to $2 \times 10^{9}$ years. The detailed equations of the code are reported in  \cite{saltelli_jasa_2002}. 

Previous works have discussed the sensitivity analysis of this simulator using alternative sampling methods and sizes. For instance, \cite{Saltelli2000} employ $ 3,084 $ simulator evaluations to obtain point estimates of the first and total order variance-based sensitivity indices. \cite{saltelli_jasa_2002} employ $ 10,000 $ simulator runs for the point estimation of first-order variance-based sensitivity indices, a second experiment with $16,384$ runs for the point of the first and total order sensitivity indices according to the design in \cite{Saltelli_cpc_2002} (no uncertainty in the estimates is provided). In \cite{ratto_cpc_2007}, stable patterns for the estimation of variance-based sensitivity measures are obtained at a cost of about $ 1,024$, after the input-output dataset has been used to train an emulator. In \cite{CastBorg12}, design based on substituted columns sampling and permuted columns sampling are used, with convergence at about $ 10^4 $ runs. \cite{Wei2014} propose a copula-based estimation methods that reduces the cost to about $1,000$ runs for point estimates, with $ 20 $ replicates for obtaining confidence intervals. \cite{PlisBorg14ESTMIM} apply a given-data design for the point estimators $\widehat{\eta}^\diamond_i$, $\widehat{\delta}^\diamond_i$ and $\widehat{\beta}^\diamond_i$ using a sample up to size $ n=5,000 $, with estimates becoming stable for $ n>1,000 $ runs.  
Thus, a sample of size $n=1,000 $ can be considered reflective of state of art for the identification of the key-uncertainty drivers of LevelE. 

\begin{figure}
	\begin{turn}{90}
		\begin{minipage}[c][\textwidth][c]{\textheight}
			\includegraphics[width=\textwidth]{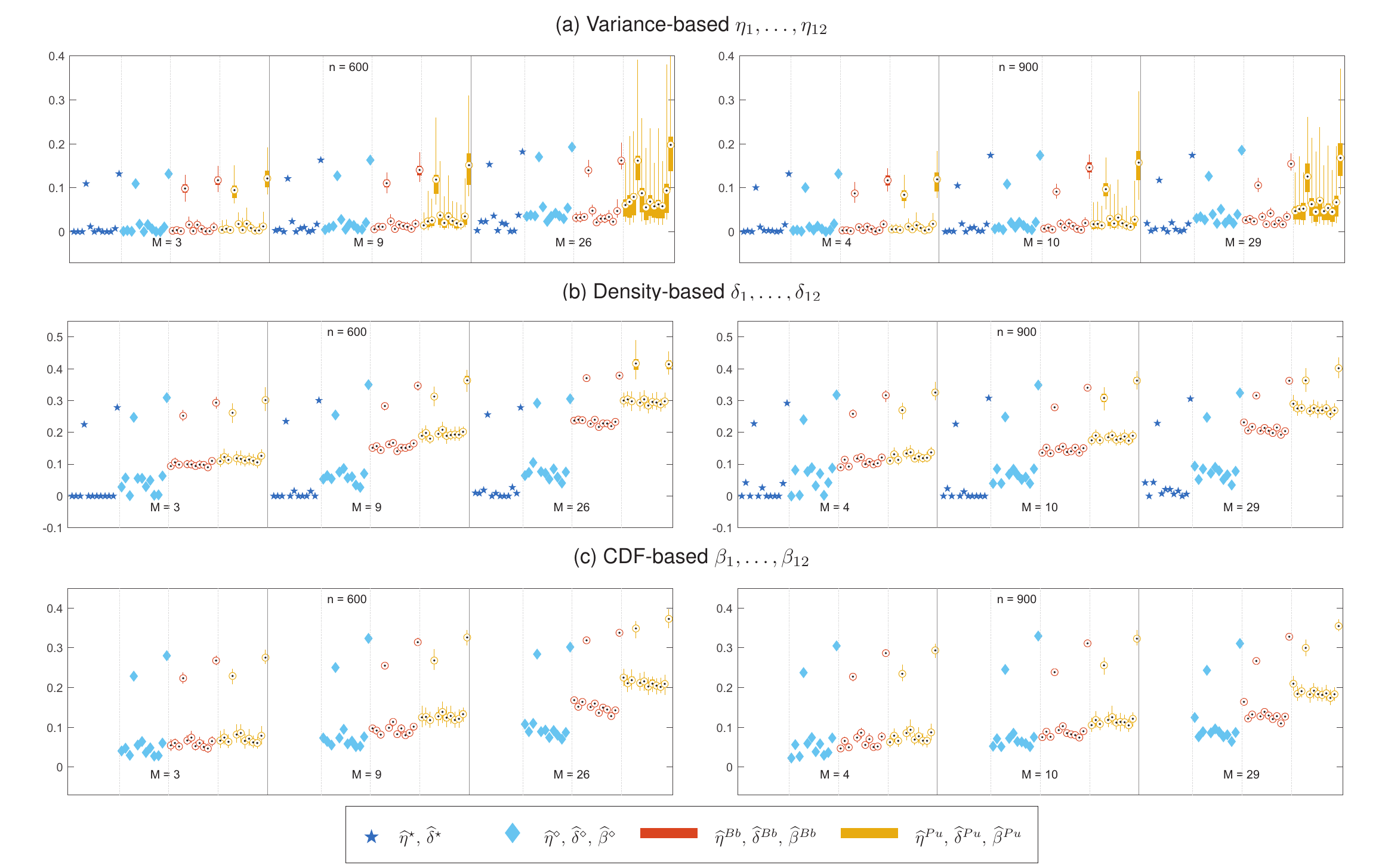}
			\caption{Results for the LevelE code: Comparison of sensitivity measures estimates using frequintist pdf/cdf-based estimators and Partition-dependent Bayesian non-parametric estimators. Bayesian estimates include 95\% credibility intervals.}
			\label{fig_BNP_LevelE_onesample}	
		\end{minipage}		
	\end{turn}
\end{figure}

\begin{figure*}[h]
	\centering
	\includegraphics[width=\textwidth]{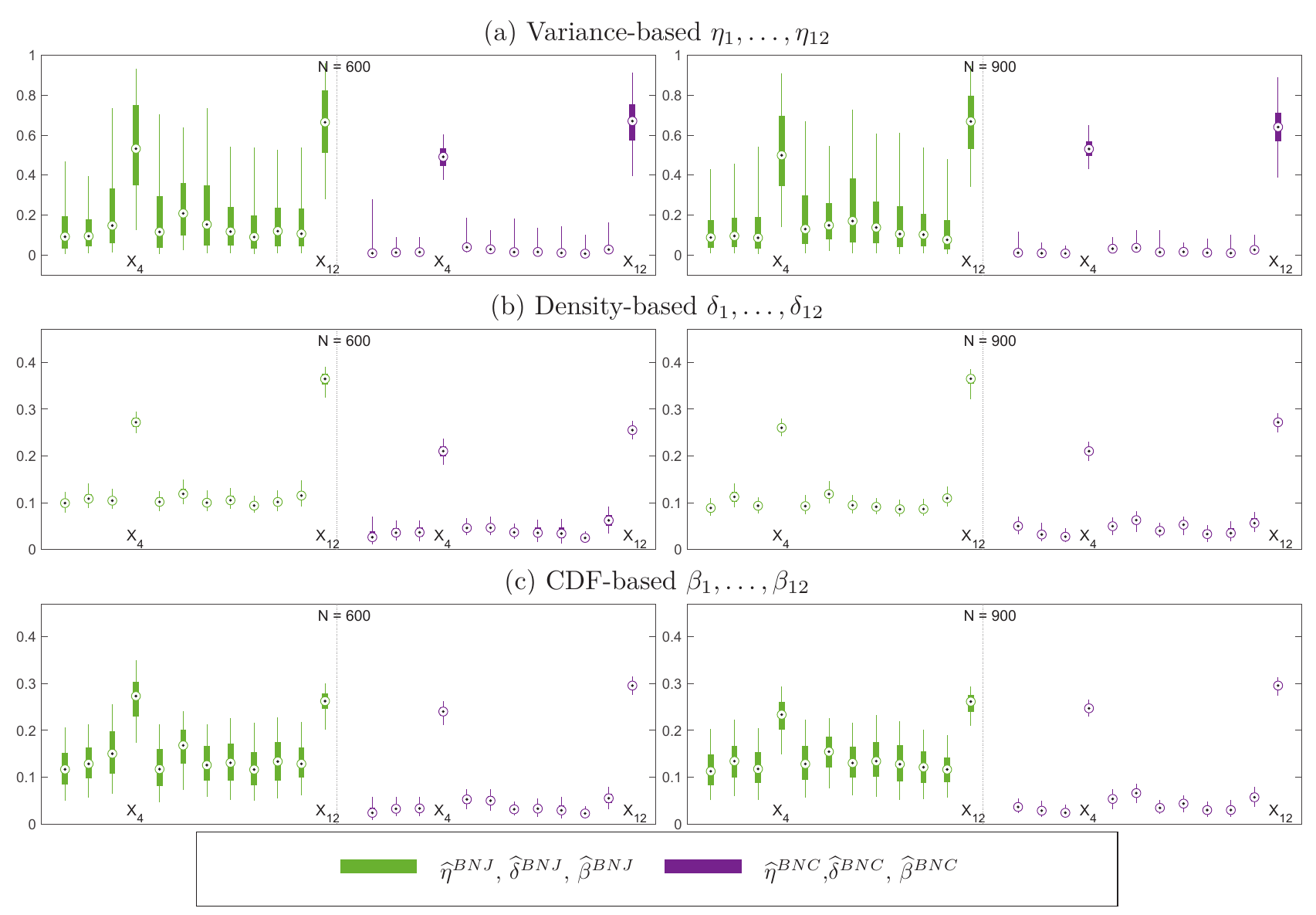}
	\caption{Results for the LevelE code: Comparison of sensitivity measures estimates with 95\% credibility intervals using Bayesian non-parametric partition-free joint/conditional estimators.}
	\label{fig_BNP_LevelE}
\end{figure*}

We report results for the calculation of global sensitivity measures using all classes of estimators discussed in the present work for samples of sizes $ n=600 $ and $ n=900 $. Figures \ref{fig_BNP_LevelE_onesample} and \ref{fig_BNP_LevelE} display the results.

The graphs in Fig. \ref{fig_BNP_LevelE_onesample} report the Bayesian bootstrap and P\'olya urn estimators, vis-\'a-vis the point estimators for variance-based (graphs in row \emph{a}), density-based (graphs in row \emph{b}) and cdf-based (graphs in row \emph{c}) sensitivity measures. The results show that already at $ n=600 $ the two most important simulator inputs are correctly identified. However, the estimates are sensitive to the partition size. Consider the right graph in row \emph{a)}. The credibility intervals of the variance-based P\'olya urn estimators with $M=26$ are completely overlapping. This signals that, had the analyst chosen such partition size, the estimates would not be meaningful. The separation becomes, instead, clearer at smaller partition sizes with $M=9$ being possibly the optimal choice. Note that the estimates tend to be upward biased as the partition size increases, in agreement with our previous experiments and also with previous literature findings.

We then come to the joint and conditional partition-independent Bayesian density estimators (Fig. \ref{fig_BNP_LevelE}). 
The two graphs in row \emph{(a)} display the estimates and credibility intervals for variance-based sensitivity measures ($ \widehat{\eta}^{BNJ}_i $,$ \widehat{\eta}^{BNC}_i $), the two graphs in row \emph{(b)} for density-based sensitivity measures ($\widehat{\delta}^{BNJ}_i $, $\widehat{\delta}^{BNC}_i $) and the two graphs in row \emph{(c)}  for cdf-based ($\widehat{\beta}^{BNJ}_i $, $ \widehat{\beta}^{BNC}_i $) sensitivity measures. 
Figure \ref{fig_BNP_LevelE} shows that the two key-uncertainty drivers are correctly identified already at $ n=600 $, by $ \widehat{\eta}^{BNC}_i $, $ \widehat{\delta}^{BNC}_i $ and $ \widehat{\beta}^{BNC}_i $, as the credibility intervals of the associated sensitivity measures separate from the credibility intervals of the remaining simulator inputs. The $ \widehat{\delta}^{BNJ}_i $ and $  \widehat{\beta}^{BNJ}_i $ correctly identify the two most influential simulator inputs.  However, $ \widehat{\eta}^{BNJ}_i $ fails to produce meaningful results for variance-based sensitivity measures at either sample sizes. This confirms the results of Section \ref{sec:simstud}. The deviation from normality strongly affects the ability of BNJ to capture the conditional density of $ Y $ given $X_i$, since much of the information contained in the data goes into the estimation of unnecessary components of the density mixture of the marginal distribution of $X_i$. This reduces the estimation precision, leading to the wider confidence intervals.

Let us consider the perspective of an analyst interpreting the results overall. From the available \emph{Data}, the analyst is able to obtain alternative estimators for representatives of three categories of sensitivity measures, with display of credibility intervals. With the exception of $ \widehat{\eta}^{BNJ}_i $, the estimators communicate that uncertainty in the simulator response is mostly driven by two simulator inputs, with the remaining ones being of lower significance. Thus, the analyst is allowed to confidently report the key-uncertainty drivers to the decision-maker even if the sample size is limited.  At the same time, Fig.s \ref{fig_BNP_LevelE_onesample} and \ref{fig_BNP_LevelE} communicate that the sample is not sufficient to rank the medium and low-important simulator inputs with confidence. If the decision-maker (modeler) wished sharper estimates of the sensitivity measures of these inputs, the analyst would need a larger sample size. This could be obtained either through additional runs of the original simulator or by fitting an emulator and, in case the fit is accurate, running the emulator instead of the original code. 

\section{Conclusions}\label{sec:conclusions}
This work has presented a fully Bayesian approach to the estimation of probabilistic sensitivity measures from a given sample. The proposed algorithms yield credibility intervals for the estimates without increasing computational burden. We have studied four classes of estimators. The first two find their theoretical ground in non-parametric Bayesian estimation based on the Dirichlet process. These estimators run in parallel with one-sample frequentist estimators currently in use, produce uncertainty in the estimates and are computationally simple to implement. However, they leave the analyst with the problem of choosing the optimal partition. The introduced conditional and unconditional non-parametric Bayesian estimators eliminate the partition selection problem, while producing uncertainty in the estimates. However, their numerical implementation needs to be carefully executed, as it requires a combination of numerical integration and MCMC. Algorithms are available, but their convergence might take a longer time than the Bayesian bootstrap and P\'olya urn estimators. Then, how should one proceed in a practical situation? The several numerical experiments performed by the authors (of which a subset was reported in the paper) evidence that the estimators succeed in identifying key-uncertainty drivers at small sample sizes in most situations. Then, a suggested approach would be to apply first the Bayesian bootstrap and/or P\'olya urn estimators on the available sample for computing an ensemble of sensitivity measures (e.g., $ \eta, \delta, \beta $). If the sensitivity measure estimates and credibility intervals yield a clear picture of the simulator inputs influence, then the analysis could be considered satisfactory.  However, the analyst ought to test this assertion repeating the estimates at alternative partition sizes. In case results are strongly dependent on the partition size, the analyst can invest in the Bayesian non-parametric estimation. If these estimators yield a clear picture about the simulator input influence, the analysis is conclusive. Conversely, a larger sample is needed and the analyst ought to plan for additional simulator runs.

While we have discussed three well-known global sensitivity measures, the paradigm presented here can be applied to the estimation of any global sensitivity measure, including, among others, value of information, sensitivity measures based on any discrepancy between densities or cumulative distribution functions.

From a more general perspective, the work shows that combining recent advances in Bayesian non-parametric density estimation with probabilistic sensitivity analysis in DACE may lead to improvements in the estimation of global sensitivity measures. Research in Bayesian non-parametric density estimation is active in Statistics and Machine Learning, but the advances in this discipline are not directly known to the DACE community. This work represents a first systematic bridge between these two closely related areas of Statistics, and we hope it could favour further research for transferring findings in Bayesian-non parametric estimation to the field of computer experiments. At the same time, exposing Bayesian estimation to the demands coming from probabilistic sensitivity analysis of realistic simulators may challenge state of the art and stimulate further research in Bayesian-non parametric estimation.

\appendix
\section{Appendix}
\subsection{Numerical experiments for the partition selection problem}\label{Sec:AppHeuristic}
The authors performed several thought experiments on test cases. The results show the difficulty, maybe impossibility, of finding a universally valid rule for linking the partition size $M$ to the sample size $n$. We report some experiments results. 

Assume the analyst wants to find an \textquotedblleft optimal \textquotedblright (in some sense) partition refining strategy, i.e., a relationship that produces the partition size $M$ that minimizes the estimation error at sample size $n$ for the pdf-based point estimators $ \widehat{\eta}_i^{\star} $, $ \widehat{\delta}_i^{\star} $ and cdf-based point estimator $\widehat{\beta}_i^{\diamond} $ [eqs. \eqref{eq_etaOS},  \eqref{eq_deltaPDF} and \eqref{eq:betaest}]. We focus on one estimator  type for simplicity and also because  \cite{BorgHazePlish15} propose an heuristic inspired by the rule of histogram partitioning of Freedman-Diaconis \citep{freedman1981histogram}, in which $M\sim \sqrt[3]{n}$. 

To evaluate the estimators' performance at fixed values of $M$ and $n$, we use the Root Mean Square Error (RMSE):
\begin{equation*}
	\text{RMSE}_i(n) \approx \sqrt{\frac{\sum_{s=1}^{S}\left(\widehat{\xi}^{i}_{s}(n)- 
			\xi_i \right)^2}{S}}
\end{equation*}
where $S$ is the number of bootstrap replicates. $\hat{\xi}_{i,l}$ is the $l$-th bootstrap replicates of $\xi_i$. 

\begin{figure*}[h]
	\vspace{-7em}
	\begin{minipage}[c][\textwidth][c]{1\textwidth} 
		\subfigure[\label{fig_heat_analy_eta} Variance-based  $\widehat{\eta}_1^{\star}$ ]{\includegraphics[width=.3\textwidth]{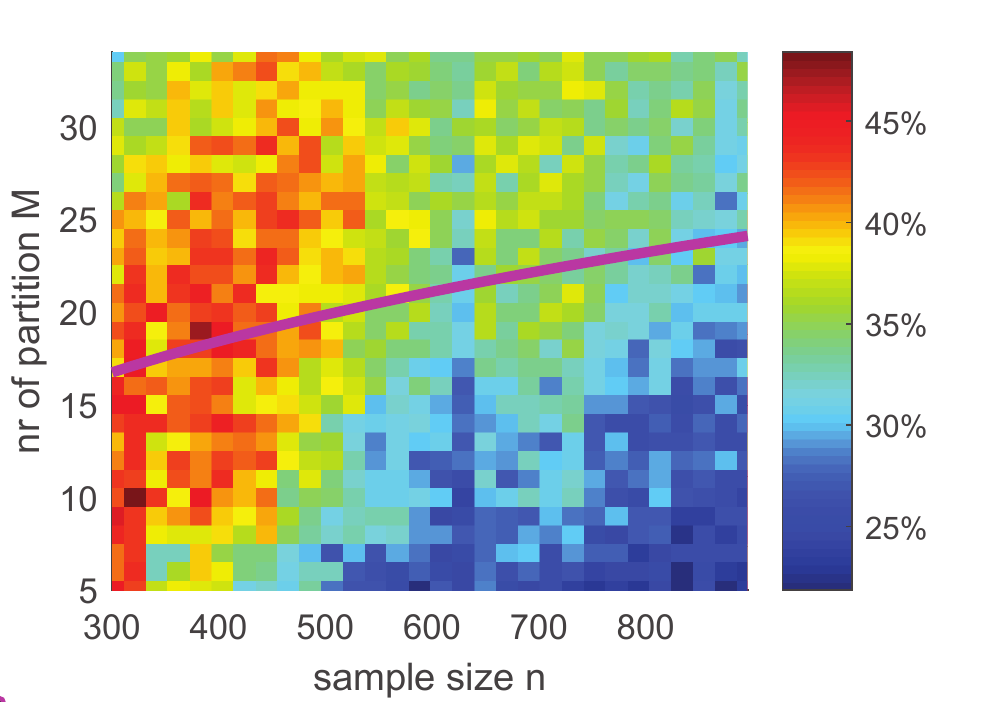}}
		\subfigure[\label{fig_heat_analy_BD} Density-based $\widehat{\delta}_1^{\star}$ ]{\includegraphics[width=.3\textwidth]{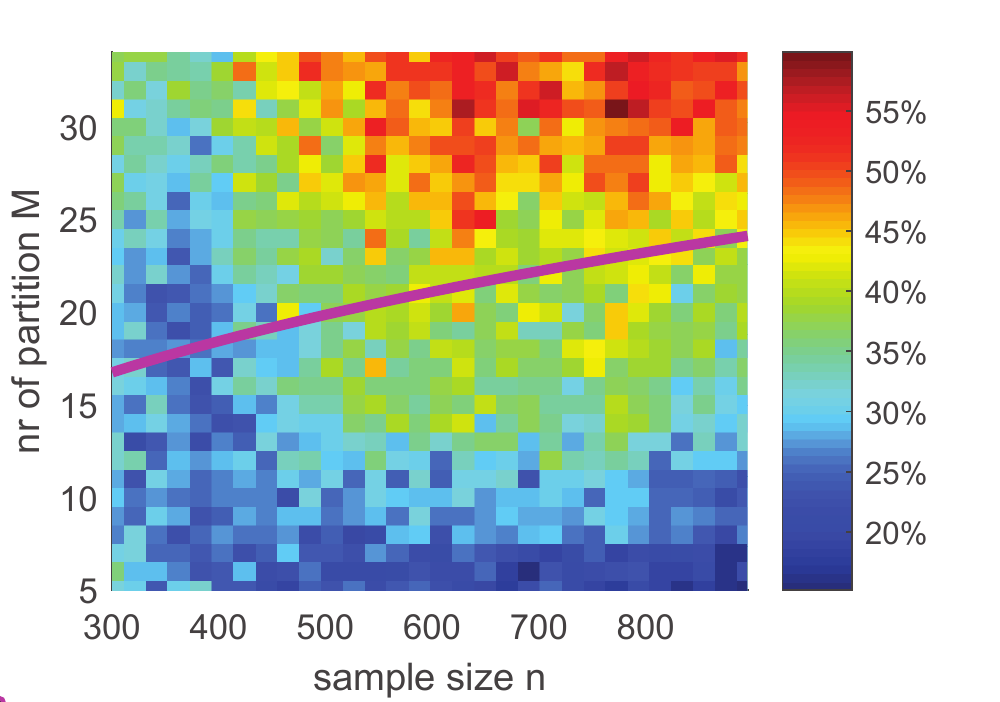}}
		\subfigure[\label{fig_heat_analy_ks} CDF-based $\widehat{\beta}_1^{\diamond}$ ]{\includegraphics[width=.3\textwidth]{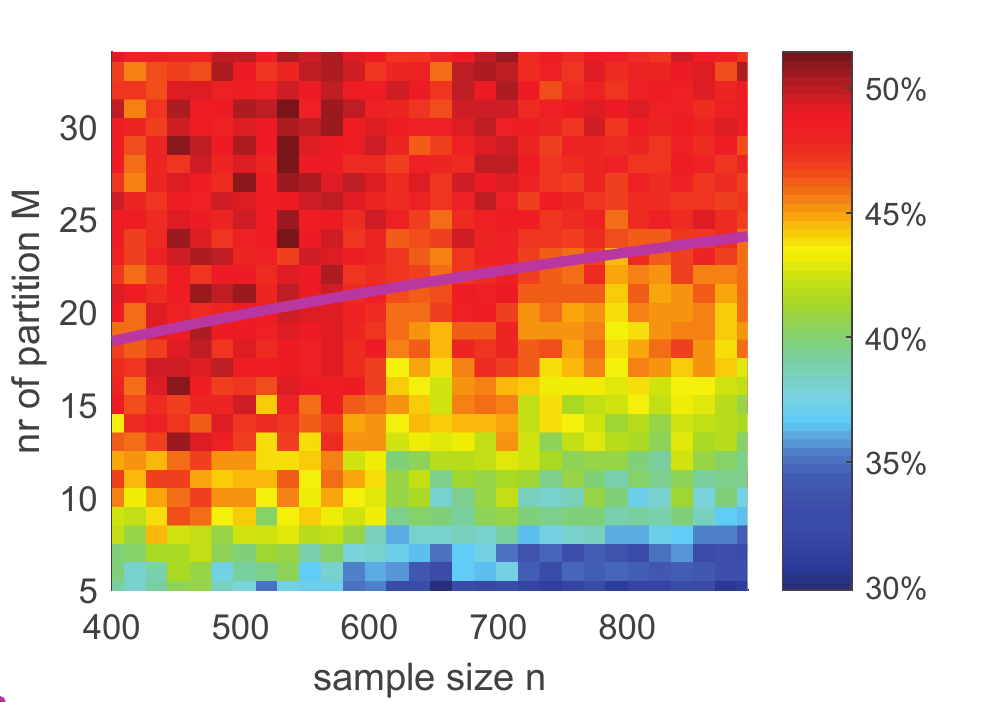}}
		\vspace{-1em}
		\caption{RMSE of sensitivity measures estimates for $X^1$ of the 2-input simulator in Eq. \eqref{eq:NLNA}. 
			Magenta lines correspond to $M=2.5\sqrt[3]{n}$; $n \in [300,900]$, $M\in [5,34]$}
		\label{fig_heatplot1}
		\vspace{-0.5em}
		\subfigure[\label{fig_heat_AG_eta} Variance-based  $\widehat{\eta}_3^{\star}$ ]{\includegraphics[width=.3\textwidth]{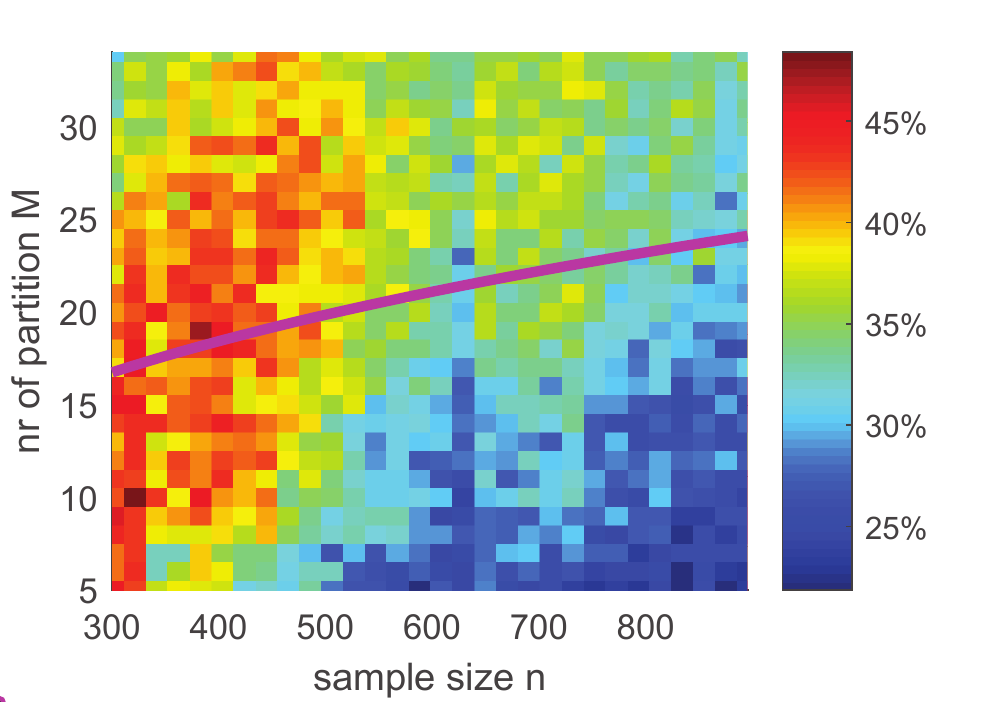}}
		\subfigure[\label{fig_heat_AG_BD} Density-based $\widehat{\delta}_3^{\star}$ ]{\includegraphics[width=.3\textwidth]{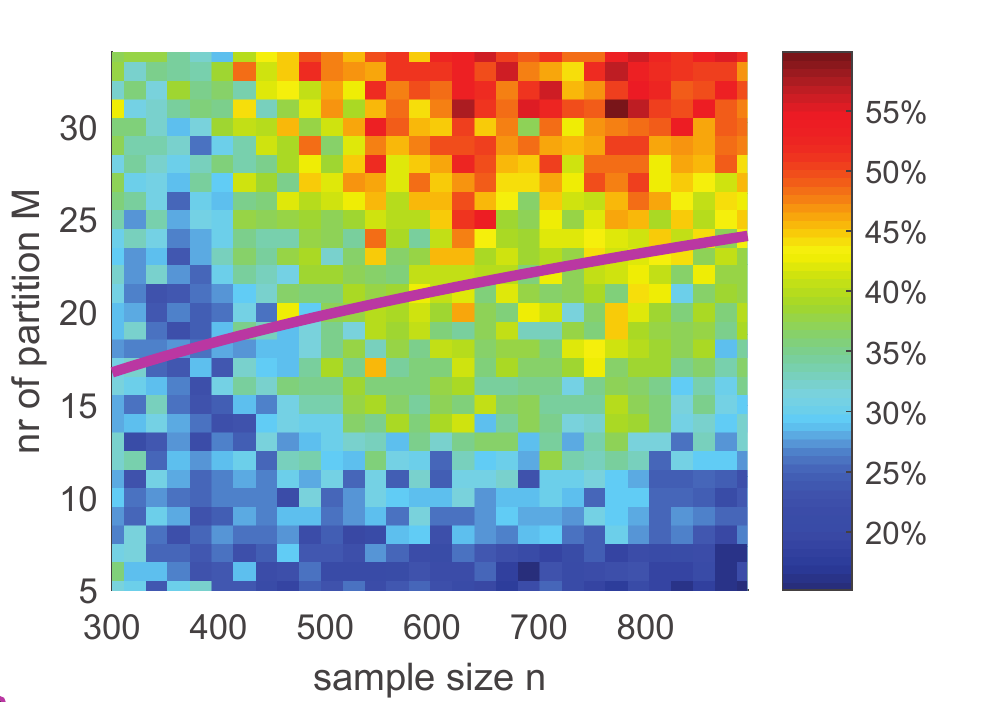}}
		\subfigure[\label{fig_heat_AG_ks} CDF-based $\widehat{\beta}_3^{\diamond}$ ]{\includegraphics[width=.3\textwidth]{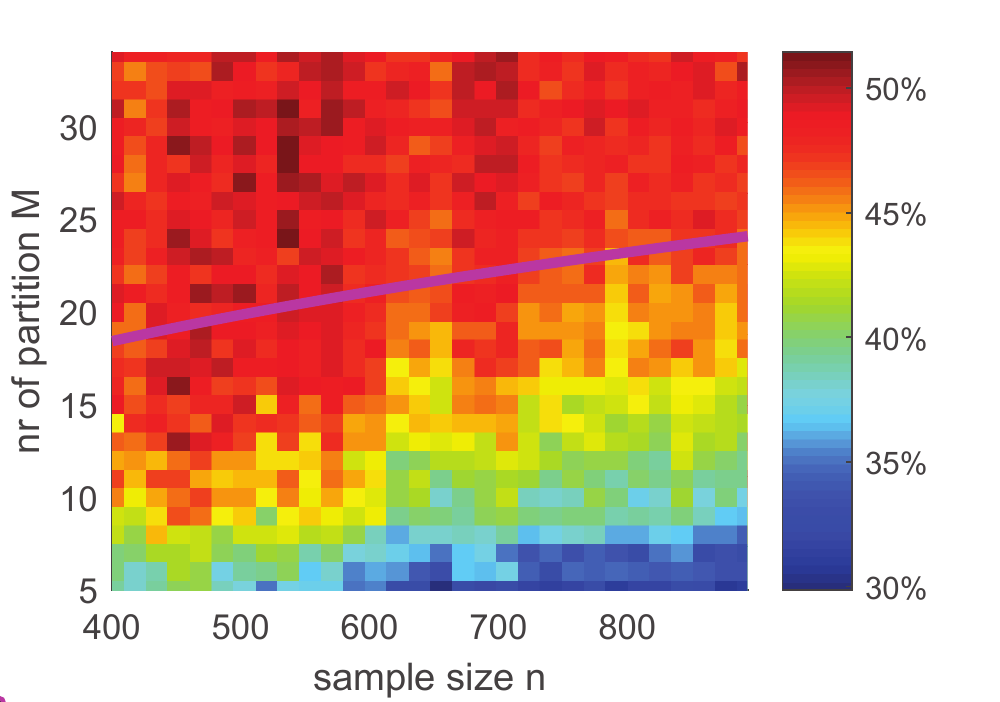}}
		\vspace{-1em}
		\caption{RMSE of sensitivity measures estimates for $X^3$ of the 21-input simulator in Eq. \eqref{eq:GauMod}. Magenta lines correspond to $M=2.5\sqrt[3]{n}$; $n \in [300,900]$, $M\in [5,34]$}
		\label{fig_heatplot2}
	\end{minipage}
	\vspace{-7em}
\end{figure*}

We estimate the sensitivity measures with sample sizes varying from $300$ to $900$, and partition sizes covering the natural numbers between $5$ and $35$. Then we calculate the RMSEs with $S=100$ bootstrap replicates. Figures \ref{fig_heatplot1} and \ref{fig_heatplot2} present the heatplot of RMSEs in percentage (RMSE$_i/\xi_i\cdot 100\%$). The horizontal axis indicates the sample size, and the vertical axis the partition size.  The darker the color of a region in the plot, the lower the estimation error. For example, in Fig. \ref{fig_heatplot1}(a), dark (blue) refers to low RMSE (less than 10 percent), and light (red) to relative high RMSE (higher than 14 percent). The magenta line maps n into M using the previously mentioned heuristic function. 
Figure  \ref{fig_heatplot1} shows that the proposed heuristic works well on the 2-input simulator (Eq. \eqref{eq:NLNA}), with the magenta line falling mainly into dark coloured regions. However, for the 21-input simulator (Eq. \eqref{eq:GauMod}) we would incur in high errors at small sample sizes. For instance consider graph a) in Fig. \ref{fig_heatplot2}. The graph reports the error in the estimates of $\delta_3$ for the second model. The heuristic would propose values of $M$ at about $20$ for all values of $n$ as optimal partition sizes. However, the partition size that minimizes the error is at about $M=10$ or lower. The different behavior here could also be related to the differences in structure and dimensionality of the models.
However, even for the same model, the heatplots differ significantly across the sensitivity measures. For the first simulator (Eq. \eqref{eq:NLNA}), the ideal partition size for the variance-based estimator is between 10 to 15 (Fig. \ref{fig_heatplot1} (b)), while for mutual information, if falls between 20 to 25 (Fig. \ref{fig_heatplot1} (c)).\\
These results show that aiming at postulating a universally valid heuristic might be a cumbersome task. 

\subsection{Implementation details for the Bayesian non-parametric estimators}\label{subsec:algos}
We present further details regarding the implementation of the Bayesian non-parametric estimation methods in Sections \ref{sec:BNPCDF} and \ref{sec:BNPPDF}. Inference on the three selected sensitivity measures $\eta_i$, $\beta_i$ and $\delta_i$ is performed independently for each $i = 1,\ldots,k$. Therefore, in order to simplify the notation, we will leave out the index $i$ throughout this appendix, considering its value fixed.
Throughout this section, all the integrals are approximated numerically using trapezoidal rule, and all the supremes are approximated by the maximum on a predetermined grid over $\mathcal{Y}$.

\subsubsection{Partition-dependent bootstrap and  P\'olya urn estimation}\label{alg_OSBB}
Recall that in Sections \ref{sec:BNPCDF}, given $M$, we have the partition $\{\mathcal{X}_m \}^M_{m=1}$ of $\mathcal{X}$ according to the sample proportion and corresponding $\{\mathbf y_m\}$. 

Within each partition set, we generate $n - n_m$ new points $\widetilde{\mathbf y}_m^{s}$ and obtain the extended vector $\mathbf{y}_m^{C,s} = (\mathbf y_m, \widetilde{\mathbf y}_m^{s})$ with $C \in \{Bb, Pu\}$, where
$\widetilde{\mathbf y}_m^{s}$ is sampled from the posterior mean $\widetilde{G}_{m}$ for $C = Bb$, and is generated through  P\'olya urn scheme when $C = Pu$. The superscript $s$ is used to indicate the $s-$th replicate.

After repeating the sampling procedure for $S$ times, we obtain the partition-depended Bayesian estimator of $\eta$ by calculating the Monte Carlo average:
\footnotesize
\begin{equation}\label{eq:etaBP}
	\widehat{\eta}^{C} = \frac{1}{S} \sum_{s=1}^{S} \eta^{C,s}, \quad \text{with } \quad \eta^{C,s} = \sum_{m=1}^{M}  \frac{n_m}{N} \frac{\left( \bar{y}_m^{C,s} - \bar{y}\right)^2  }{s_y^2} ,
\end{equation}
\normalsize
where $\bar{y}_m^{C,s}$ is the sample mean of $\mathbf{y}_m^{C,s}$; $\bar{y}$ and $s_y^2$ are the sample mean and variance of $\mathbf{y}$. Approximate credibility intervals of $\eta$ can be obtained from the empirical quantiles of $\{\eta^{C,s}, s=1\dots,S\}$. The same intuition is used for $\delta$ and $\beta$. Specifically, we use
\footnotesize
\begin{eqnarray}
	\widehat{\delta}^{C} = \frac{1}{S} \sum_{s=1}^{S} \delta^{C,s}, \quad \text{with } \quad \delta^{C,s} = \sum_{m=1}^{M} \frac{n_m}{N}  \int_{\mathcal{Y}}|\hat f^{\star}_{Y}(y)-\hat f_m^{C,s} (y)|\text{d}y , \label{eq:deltaBP}\\
	\widehat{\beta}^{C} = \frac{1}{S} \sum_{s=1}^{S} \beta^{C,s}, \quad \text{with } \quad \beta^{C,s} = \sum_{m=1}^{M} \frac{n_m}{N}  \sup_{y \in \mathbf{y}_m^{C,s}} \left\vert \hat F_Y(y)-\hat F_m^{C,s}(y) \right\vert, \label{eq:betaBP}
\end{eqnarray}
\normalsize
where $ \hat{f}^{\star}_{Y} $ and $ \hat f_m^{C,s}$ are kernel smoothing functions of $\mathbf{y}$ and  $\mathbf{y}_{m}^{C,s}$, respectively; $\hat F_Y$, and $\hat F_m^{C,s}$ are the empirical cdf's of $\mathbf{y} $ and $\mathbf{y}_m^{C,s}$, respectively.

Note that the calculations of $\eta^{C,s}$, $\delta^{C,s}$ and $\beta^{C,s}$ are equivalent to the pdf-based estimators in eqs. \eqref{eq_etaOS}, \eqref{eq_deltaPDF}, \eqref{eq:betaest} but with the enriched samples.
Alternatively, the cdf-based estimators in eqs. \eqref{eq:etafromcdf} and \eqref{eq:deltafromcdf} could be used for $\eta^{C,s}$ and $\delta^{C,s}$. 

\subsubsection{Partition-free joint density-based estimation}	\label{alg_BNJ}
Following the proposal in \citet{jara2011dppackage}, we fix $\alpha = 1$, and choose $G$ to be a Normal-Inverse Wishart distribution
\footnotesize
\begin{equation*}
	(\mu_\ell, \Sigma_\ell)| (m_1, \gamma, \psi_1)\iid \mathcal{N}(\mu_{\ell} \vert m_1, \frac{1}{\gamma} \Sigma) IW(\Sigma_{\ell} \vert
	4, \psi_1),\quad \ell=1,2,\ldots,
\end{equation*}
\normalsize
where $\mathcal{N}(\cdot|m,A)$ denotes a bivariate normal distribution with mean $m$ and covariance matrix $A$, and $IW(\cdot |4, \psi)$ denotes an Inverse-Wishart distribution with mean $\psi^{-1}$. A hyper-prior is assigned to the parameters of the base measure, with hyperparameters determined empirically:
\footnotesize
\begin{equation*}
	\gamma \sim \text{Gamma} \left(\cdot|0.5, 0.5\right),\;
	m_1 \vert (m_2, s_2) \sim \mathcal{N}(\cdot|m_2, s_2),\;
	\psi_1 \vert (s_2) \sim IW (\cdot|4, s_2^{-1}),
\end{equation*}
\normalsize
where $\text{Gamma} (\cdot|a_1,a_2)$ denotes the  Gamma distribution with mean $a_1/a_2$; $m_2 = (\mu_{X}, \bar y)$ and $s_2 = \text{diag}(\sigma_{X}^2, s_y^2)$.

Inference is achieved through the function \texttt{DPdensity} from the \texttt{DPpackage} in \texttt{R}. The output is a MCMC posterior sample $\underline{\theta}^s = (\underline{w}^s,\underline{\mu}^s,\underline{\Sigma}^s)$,  $s=1,\ldots,S$. In practice, the number $J_s$ of components with non-zero weights is finite, thus we have
\footnotesize
\begin{eqnarray}
	\underline{w}^s = (w_1^s, \dots, w_{J_s}^s), \quad
	\underline{\mu}^s &=& (\mu_1^s, \dots, \mu_{J_s}^s), \quad
	\underline{\Sigma}^s = (\Sigma_1^s, \dots, \Sigma_{J_s}^s),
	\notag\\ 
	\text{with} \qquad
	\mu_{\ell}^s &=& \left[ {\begin{array}{c}
			\mu_{1,\ell}^s \\
			\mu_{2,\ell}^s \\
	\end{array} } \right], \quad
	\Sigma_{\ell}^s  = \left[ {\begin{array}{cc}
			\sigma_{1,\ell}^s & \sigma_{3,\ell}^s \\
			\sigma_{3,\ell}^s & \sigma_{2,\ell}^s\\
	\end{array} } \right] .
\end{eqnarray}
\normalsize
Given the posterior realizations, the corresponding joint density can be obtained:
\footnotesize
\begin{equation*}
	f^{BNJ,s}_{X,Y} (x,y|\underline{\theta}^s) = \sum_{\ell=1}^{J_s} w_\ell^s \cdot \mathcal{N}
	(x,y|\mu_\ell^s,\Sigma_\ell^s).
\end{equation*}
\normalsize
By the properties of the bivariate Normal distribution, the marginal and conditional distributions, $f^{BNJ,s}_{Y}$ and $f^{BNJ,s}_{Y\vert X^i}$ recpectively, are also mixtures of Normal distributions:
\footnotesize
\begin{equation} \label{eq:jointmarginalfY}
	f^{BNJ,s}_{Y} (y \vert \underline{\theta}^s) = \sum_{\ell=1}^{J_s} w_\ell^s \cdot \mathcal{N}
	(y|\mu_{2,\ell}^s,\sigma_{2,\ell}^s),\quad
	f^{BNJ,s}_{Y\vert x} (y \vert x,\underline{\theta}^s )=  \sum_{\ell=1}^{J_s} w_\ell^s \cdot \mathcal{N} \left(\cdot|\nu_{2,\ell}^s,\tau_{2,\ell}^s \right)
\end{equation}
\normalsize
where $\nu_{\ell}^s = \mu_{2,\ell}^s + \sigma_{3,\ell}^s (x - \mu_{1,\ell}^s )/\sigma_{1,\ell}^s$ and $\tau_{\ell}^s = \sigma_{2,\ell}^s - (\sigma_{3,\ell}^s )^2/\sigma_{1,\ell}^s$. Clearly, the corresponding cdfs, $F^{BNJ,s}_{Y}$ and $F^{BNJ,s}_{Y \vert X}$, as well as the marginal mean and variance can be calculated trivially. In particular, 
\footnotesize
\begin{equation}\label{eq:joitmarginalY}
	\mu_{Y}^s \coloneqq \mathbb{E}[Y \vert \underline{\theta}^s]
	= \sum_{\ell=1}^{J_s} w_{\ell}^s \mu_{2,\ell}^s, \quad
	V_Y^s \coloneqq \mathbb{V}[Y \vert \underline{\theta}^s] =  \sum_{\ell=1}^{J_s} w_{\ell}^s \left(\sigma_{2,\ell}^s + \left(\mu_{Y}^s - \mu_{2,\ell}^s \right)^2 \right) .
\end{equation}
\normalsize

Thus, MCMC samples of the sensitivity measures of interest can be obtained as follows:
\footnotesize
\begin{eqnarray*}
	& &\eta^{BNJ,s} \approx \frac{V^s}{V_Y^s};\quad 
	\delta^{BNJ,s} \approx \frac{1}{2}\int_{\mathcal{X}}\int_{\mathcal{Y}}\left\vert 
	f^{BNJ,s}_{X, Y}-f_{X}\cdot f^{BNJ,s}_{Y}\right\vert \text{d}y\text{d}x;\\
	& &\beta^{BNJ,s}\approx\int_{\mathcal{X}} \sup_{\mathcal{Y}} \left| F^{BNJ,s}_Y - F^{BNJ,s}_{Y|X} \right| f_{X} \text{d}x,
\end{eqnarray*}
\normalsize
where
\footnotesize
\begin{equation*}
	\mu_{Y}^s(x) \coloneqq \mathbb{E}[Y \vert X=x, \underline{\theta}^s] = \sum_{\ell=1}^{J_s} w_\ell^s \nu_{2,\ell}^s,
\end{equation*}
\begin{equation*}
	V^s = \int_{\mathcal{X}} \left(\mu_{Y}^s(x) - \mu_Y^s  \right)^2 f_{X} dx = 
	\int_{\mathcal{X}} \left( \sum_{\ell=1}^{J_s} w_{\ell}^s \frac{\sigma_{3,\ell}^s}{\sigma_{1,\ell}^s} \left(x - \mu_{1,\ell}^s \right)  \right) ^2  f_{X} \text{d}x .
\end{equation*}
\normalsize

Point estimators of interest are obtained as Monte Carlo averages:
\footnotesize 
\begin{equation}\label{eq_BNJ}
	\widehat{\eta}^{BNJ} = \frac{1}{S} \sum_{s=1}^{S} \eta^{BNJ,s}, \qquad 
	\widehat{\delta}^{BNJ} = \frac{1}{S} \sum_{s=1}^{S} \delta^{BNJ,s}, \qquad 
	\widehat{\beta}^{BNJ} = \frac{1}{S} \sum_{s=1}^{S} \beta^{BNJ,s} .
\end{equation}
\normalsize

\subsubsection{Partition-free conditional density-based estimation} \label{alg_BNC}
\begin{figure}[!]
	\centering
	\includegraphics[width=0.5\textwidth]{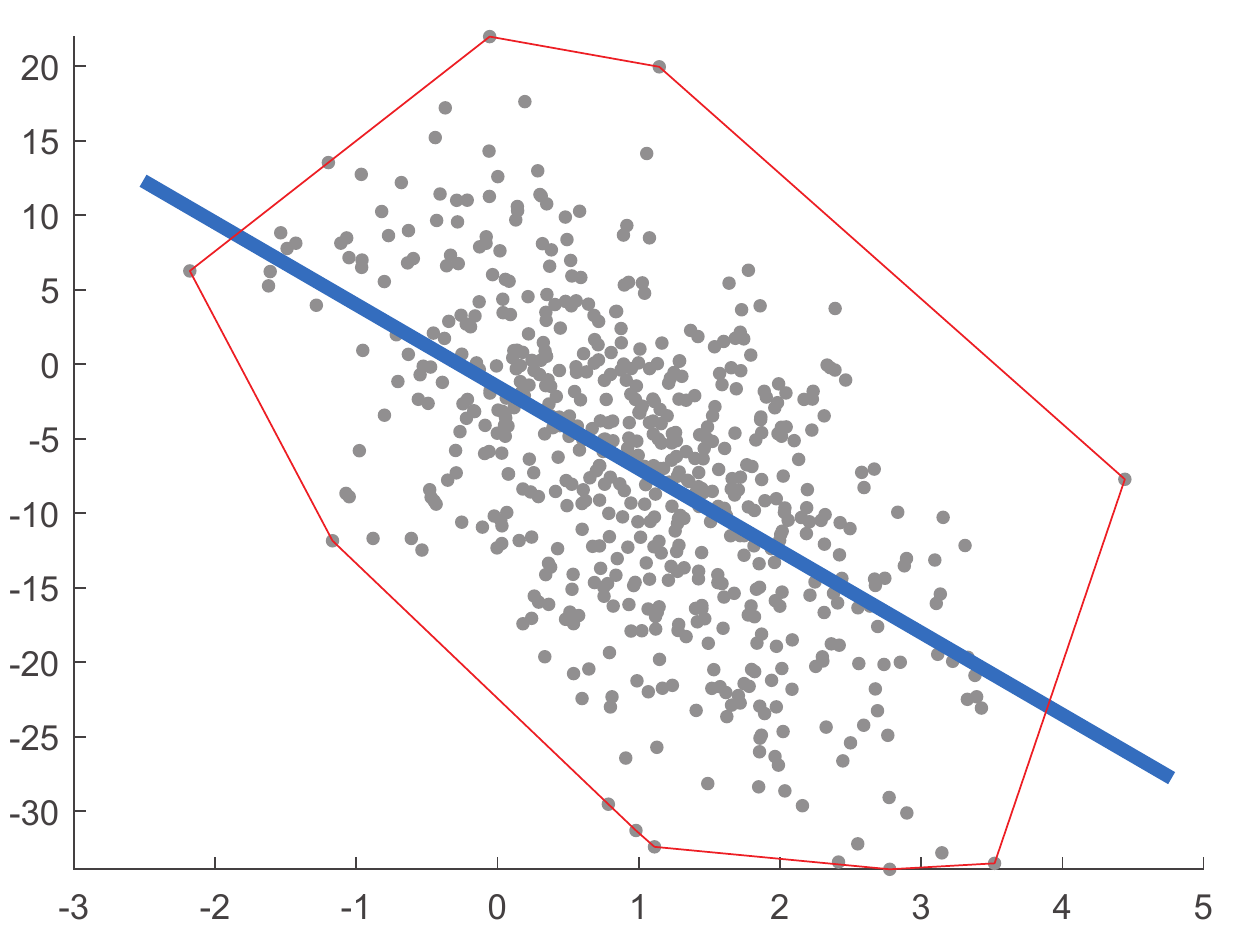} 
	\caption{Scatter plot of $Y$ and $X^3$ for the 21-input simulator. Colorful lines constitute the convex hull of $Data$. The bold red and blue lines are used for prior specification. } 
	\label{fig:choosebeta} 
\end{figure}

Following the proposal of \cite{antoniano2014bayesian}, we fix $\alpha=1$ and choose $\mathcal{K}(x|\psi_\ell)$ to be a Normal kernel,  with $ \psi_\ell =(\mu_{\ell}, \tau)$. The base measure $G$ is given by:
\footnotesize
\begin{equation*}\label{eq:isaprior}
	\tau\sim \text{Gamma} (\cdot \mid 1,1);\quad (\mathbf{b}_\ell,\sigma_{\ell},\mu_\ell)\iid \mathcal{N}\left(\mathbf{b}_\ell \mid \mathbf{b}_0 , \sigma_{\ell} C^{-1} \right) \text{Gamma} (\sigma_{\ell}^{-1}\mid 1,1) \mathcal{N} \left(\mu_\ell \mid \mu_0, \left(\tau/10 \right)^{-1}  \right),
\end{equation*}
\normalsize
where $\mathbf{b}_\ell = (a_\ell,b_\ell)$.The hyperparameters are chosen empirically. As an illustration, consider the 21-input simulator. Figure \ref{fig:choosebeta} shows the scatter-plot of $(\mathbf{x}^3,\mathbf{y})$ and the convex hull, i.e. the smallest convex set containing all points. In this case, we fix $\mathbf{b}_0 = (-1.5, -5.5)$ and $C^{-1} = \text{diag}(43^2, 11^2)$, in order to allow each local linear component to lie between the blue and red lines in the figure, which represent the main behaviour of the data.
We use the \textsc{Matlab} subroutine provided by \cite{antoniano2014bayesian} to generate an MCMC posterior sample $(\underline{\theta}^s,\underline{\psi}^s)=(\underline{a}^s, \underline{b}^s, \underline{\sigma}^s, \underline{\omega}^s, \underline{\mu}^s, \tau^s)$, $s=1\ldots S$, where
\footnotesize
\begin{eqnarray}
	\underline{a}^s &=& (a_1^s, \dots, a_{J_s}^s), \quad
	\underline{b}^s = (b_1^s, \dots, b_{J_s}^s), \quad
	\underline{\sigma}^s = (\sigma_1^s, \dots, \sigma_{J_s}^s),\notag\\ 
	\underline{\omega}^s &=& (\omega_1^s, \dots, \omega_{J_s}^s),  \quad
	\underline{\mu}^s = (\mu_1^s, \dots, \mu_{J_s}^s).
\end{eqnarray}
\normalsize
Given the a posterior realization $(\underline{\theta}^s,\underline{\psi}^s))$, a conditional density can be obtained from eqs. \eqref{eq:weightsIsa} and \eqref{eq:mixcondIsa}:
\footnotesize
\begin{equation}
	f^{BNC,s}_{Y|X} (y|x,\underline{\theta}^s,\underline{\psi}^s) = \sum_{\ell=1}^{J_s} w_\ell^s(x) \mathcal{N}\left(y|a_\ell^s+b_\ell^s  x, \sigma_\ell^{s} \right).
\end{equation}
\normalsize
The corresponding marginal pdf $f^{BNC,s}_{Y}$  of $Y$ is obtained by integrating with respect to the true $f_X$:
\footnotesize
\begin{equation}
	f^{BNC,s}_{Y}(y \vert \underline{\theta}^s) \approx \int_{\mathcal{X}} f^{BNC,s}_{Y|X} f_{X} \text{d}x.
\end{equation}
\normalsize
Clearly, the corresponding marginal and conditional cdfs, $ F^{BNC,s}_{Y|X}$ and $F^{BNC,s}_{Y}$, respectively can be obtained trivially.
In particular, posterior realizations of the marginal mean and variance of $Y$ are given by
\footnotesize
\begin{equation}
	\mu_Y^s \coloneqq \mathbb{E}[Y\vert \underline{\theta}^s,\underline{\psi}^s] \approx \int_{\mathcal{Y}} y f^{BNC,s}_{Y} \text{d}y, \quad
	V_{Y}^s \coloneqq \mathbb{V}[Y \vert \underline{\theta}^s,\underline{\psi}^s] \approx \int_{\mathcal{Y}} \left( y - \mu_Y^s \right)^2 f^{BNC,s}_{Y} \text{d}y
\end{equation}
\normalsize

Thus, MCMC samples of the sensitivity measures of interest can be obtained as follows:
\footnotesize
\begin{eqnarray*}
	&&\eta^{BNC,s} \approx \frac{V^s}{V_Y^s};\quad \delta^{BNC,s} \approx \frac{1}{2} \int_{\mathcal{X}} \int_{\mathcal{Y}} \left| f^{BNC,s}_Y - f^{BNC,s}_{Y|X} \right| \text{d}y f_{X} \text{d}x;\\
	&&\beta^{BNC,s} \approx\int_{\mathcal{X}} \sup_{\mathcal{Y}} \left| F^{BNC,s}_Y - F^{BNC,s}_{Y|X} \right| f_{X} \text{d}x,
\end{eqnarray*}
\normalsize
where
\footnotesize
\begin{equation}
	\mu_{Y}^s(x) \coloneqq \mathbb{E}[Y \vert x, \underline{\theta}^s, \underline{\psi}^s]  = \sum_{\ell=1}^{J_s} \omega_{\ell}^s (x) \left(a_{\ell} + b_{\ell}x \right).
\end{equation}
\begin{equation}
	\widetilde{\mu}_Y^s \coloneqq \mathbb{E}[\mu_{Y}^s(X)
	] \approx \int_{\mathcal{X}} \mu_{Y}^s(x) f_{X} \text{d}x, \quad
	V^s=\mathbb{V}[\mu_{Y}^s(X)]\approx \int_{\mathcal{X}} \left(\mu_{Y}^s(x) - \widetilde{\mu}_Y^s  \right)^2 f_{X} \text{d}x .
\end{equation}
\normalsize
Finally, point estimators of interest are obtained as Monte Carlo averages: 
\footnotesize
\begin{equation}\label{eq_BNC}
	\widehat{\eta}^{BNC} = \frac{1}{S} \sum_{s=1}^{S} \eta^{BNC,s}, \qquad 
	\widehat{\delta}^{BNC} = \frac{1}{S} \sum_{s=1}^{S} \delta^{BNC,s}, \qquad 
	\widehat{\beta}^{BNC} = \frac{1}{S} \sum_{s=1}^{S} \beta^{BNC,s} .
\end{equation}
\normalsize


\begin{footnotesize}
	\noindent\textbf{Note }
	The code can be downloaded from:\\
	\url{https://github.com/LuXuefei/Nonparametric-estimation-of-probabilistic-sensitivity-measures},
	along with the simulated data to reproduce results.
\end{footnotesize}

\clearpage
\bibliographystyle{plainnat}

\end{document}